\documentstyle [11pt,aaspp4]{article}
\begin{document}
\input{epsf}
\newcommand{\rot}{\mathop{\rm rot}\nolimits}
\newcommand{\erg}{\mathop{\rm erg}\nolimits}
\newcommand{\grad}{\mathop{\rm grad}\nolimits}
\newcommand{\diver}{\mathop{\rm div}\nolimits}
\newcommand{\const}{\mathop{\rm const}\nolimits}
\newcommand{\Ma}{\mathop{\rm Ma}\nolimits}
\newcommand{\tg}{\mathop{\rm tg}\nolimits}
\normalsize

\title{SPREAD OF MATTER OVER A NEUTRON-STAR SURFACE DURING~ DISK ACCRETION}

\author{N. A. Inogamov$^{1,2}$ and R. A. Sunyaev$^{2,3}$}

\affil{$^1$ L.D. Landau Institute for Theoretical Physics,
Russian Academy of Sciences,
142432~ Chernogolovka, Moscow Region, Russian Federation\\
$^2$ Max-Planck-Institut fuer Astrophysik,
Karl Schwarzschild Strasse 1,
86740~ Garching~ bei~ Muenchen,~ Germany\\
$^3$ Space Research Institute, Russian Academy of Sciences,\\
117810~ Moscow, Russian Federation}

$^*$ To appear in Astron. Lett., 1999, vol. 25, no. 5 (May issue).

Submitted 1998, October 15.

\begin{abstract}
Disk accretion onto a slowly rotating neutron star
with a weak magnetic field $H < 3\times 10^8$ gauss
is considered
in a wide range of luminosities $1/100 < L/L_{edd} < 1,$
where $L_{edd}$ is the Eddington luminosity.
We construct a theory for the deceleration of rotation
and the spread of matter over the stellar surface
in the shallow-water approximation.
The rotation slows down due to friction
against the dense underlying layers.
The deceleration of Keplerian rotation
and the energy release take place on the stellar surface
in a latitudinal belt
whose meridional width rises with increasing $L.$
The combined effect
of centrifugal force and radiation pressure
gives rise to two latitudinal rings of enhanced brightness
which are symmetric around the equator
in the upper and lower hemispheres.
They lie near the edges
of differentially rotating and radiating
upper and lower belts.
The bright rings shift from the equatorial zone to higher latitudes
when the luminosity $L$ rises.
The ring zones are characterized by a minimum surface density
and, accordingly, by a maximum meridional spread velocity.
At a low accretion rate and luminosity,
the released energy is removed through the comptonization 
of low-frequency photons.
\end{abstract}

\section{OVERALL PICTURE}

In the Newtonian approximation,
half of the gravitational energy
is released in an extended disk during accretion onto a neutron star
with a weak magnetic field.
If the star rotates slowly,
the other half of the energy
must be released in the immediate vicinity of the star
as the matter transfers from the disk onto the surface.
This energy release is generally thought to occur
in a narrow
(both in radius and in latitude)
boundary layer of the disk
which is formed near the contact boundary
between the disk and the star.
The boundary layer is significant
not only due to contribution to the luminosity of the compact object.
It is important since the boundary-layer area is small
compared to that of the disk,
the layer's emission is harder
and must be observed in a different spectral range
relative to the disk emission.

Accretion onto a neutron star differs in many ways
from accretion onto a white dwarf.
First, the percentage of energy
which is released in the boundary layer
can exceed appreciably $50\%$
because of the general-relativity effects (Sunyaev and Shakura 1986).
The boundary-layer luminosity $L_{SL}$
is greater than the disk luminosity $L_d$ 
even for rapidly rotating neutron stars,
but the ratio $L_{SL}/L_d$ decreases
with the increasing angular velocity of the star (Sibgatullin and Sunyaev 1998).
Second, at luminosities
$L > L_{edd}/100 \sim 10^{36}$ erg s$^{-1},$
the local flux of radiative energy
which is emitted from the boundary layer
is of the order of the Eddington flux,
i.e., the force of radiation pressure on an electron
is of the order of the gravitational force acted on a proton.
There are also other dissimilarities.

Most of the authors
who considered disk accretion onto stars
(Shakura and Sunyaev 1973;
Lynden-Bell and Pringle 1974;
Pringle and Savonije 1979;
Papaloizou and Stanley 1986;
Popham et al. 1993),
white dwarfs
(Tylenda 1981;
Meyer and Meyer-Hofmeister 1989;
Popham and Narayan 1995),
and neutron stars
(Shakura and Sunyaev 1988; Bisnovatyi-Kogan 1994)
suggest that the accreting-plasma velocity
decreases from the Keplerian velocity
to the stellar rotation velocity
through the turbulent friction
between the differentially rotating layers
of accreting matter
within a thin boundary layer of the disk (see Fig. 1).
In this case, $d\omega/dr > 0$ in the boundary layer,
the boundary-layer meridional extent is
$R\theta_{bl}= R \sqrt{2 T_{bl}/m_p}/v_{\varphi}^k,$
and its height is
$h_{bl}\simeq R (2T_{bl}/m_p)/(v_{\varphi}^k)^2,$
where $R$ is the stellar radius,
$T_{bl}$ is the boundary-layer temperature
given in energy units,
$m_p$ is the proton mass,
and $v_{\varphi}^k$ is the Keplerian velocity at the radius $R$
(Shakura and Sunyaev 1988).
The meaning of the geometrical quantities $h_{bl}$ and $\theta_{bl}$
is clear from Fig. 1.
We see that in this approach,
$h_{bl}$ is determined by the scale height
in a gravity field
with a free fall acceleration
$g_0 = \gamma M/R^2 = $ $(v_{\varphi}^k)^2/R.$
The boundary-layer meridional extent, $R\theta_{bl},$
is given by the same formula that is used to calculate the disk thickness,
i.e., only the tangential component of the gravity force
 is taken into account.
The thickness, $h_{bl},$ and the meridional extent turn out to be independent
of latitude and accretion rate, respectively.

\begin{figure}
\epsfxsize=17cm
\epsffile{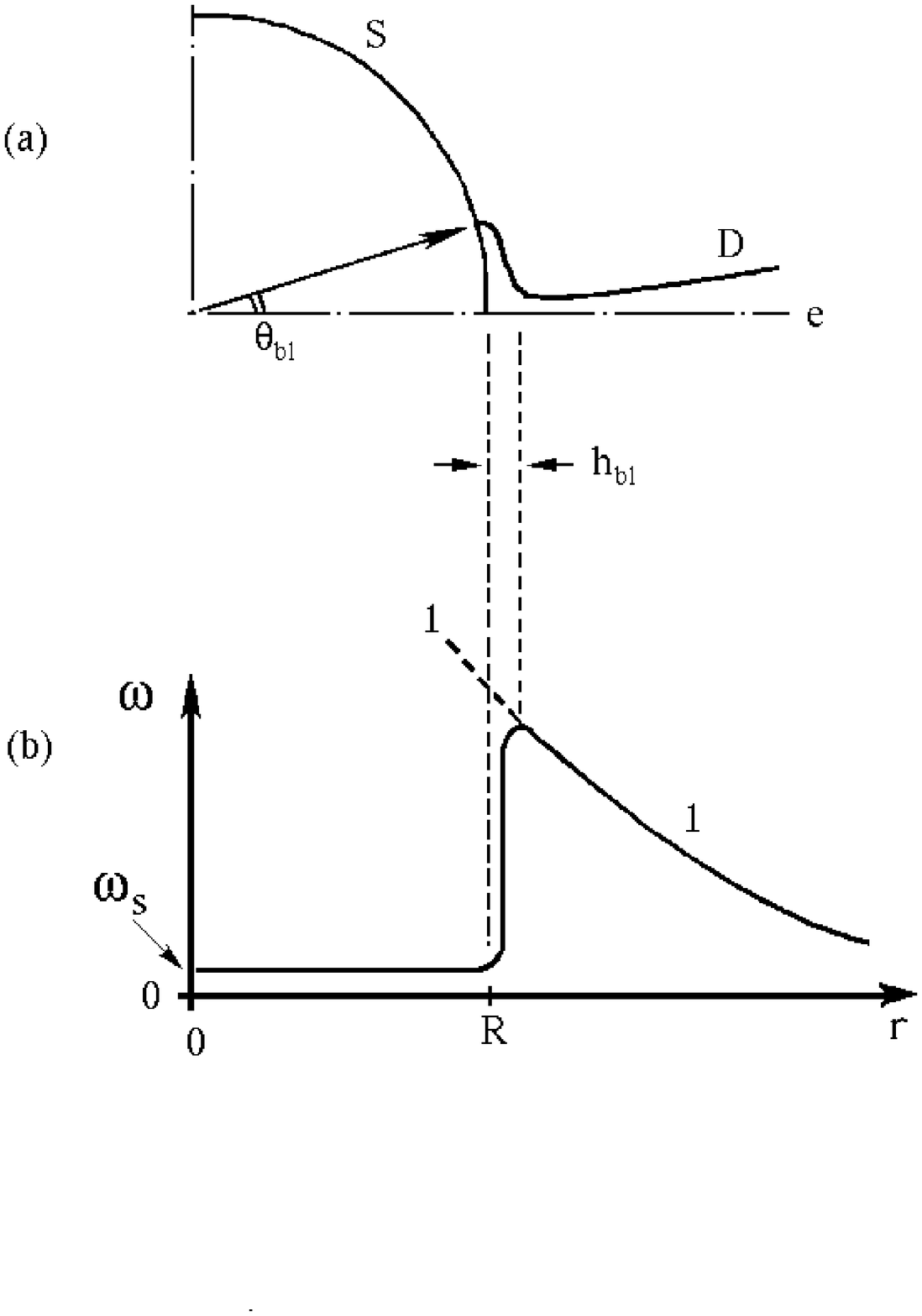}
\caption{(a) Shape of the boundary layer in the standard approach.
The notation is as follows:
$e$ is the equatorial plane,
$D$ is the disk,
and $S$ is the stellar surface.
(b) Angular velocity $\omega = v_{\varphi}/r$ versus radius,
where $v_{\varphi}$ is the rotation velocity;
1 -- the Keplerian dependence:
$\omega\propto r^{ - 3/2};$
the angular velocity $\omega_n$ in the neck
is at a maximum
and approximately equal to the Keplerian velocity,
$f_n =$
$\omega_n/2\pi \approx$ 
$ v_{\varphi}^k / 2 \pi R =$
$1.84 (M/M_{\odot})^{1/2}/R_6^{3/2}$ kHz,
$R_6 = R/(10$ km);
inside the star from the center up to its surface $r \leq R,$
the rotation is rigid: $\omega \equiv \omega_S.$}
\end{figure}

\begin{figure}[tbh]
\epsfxsize=10cm
\epsfbox[-300 87 400 753]{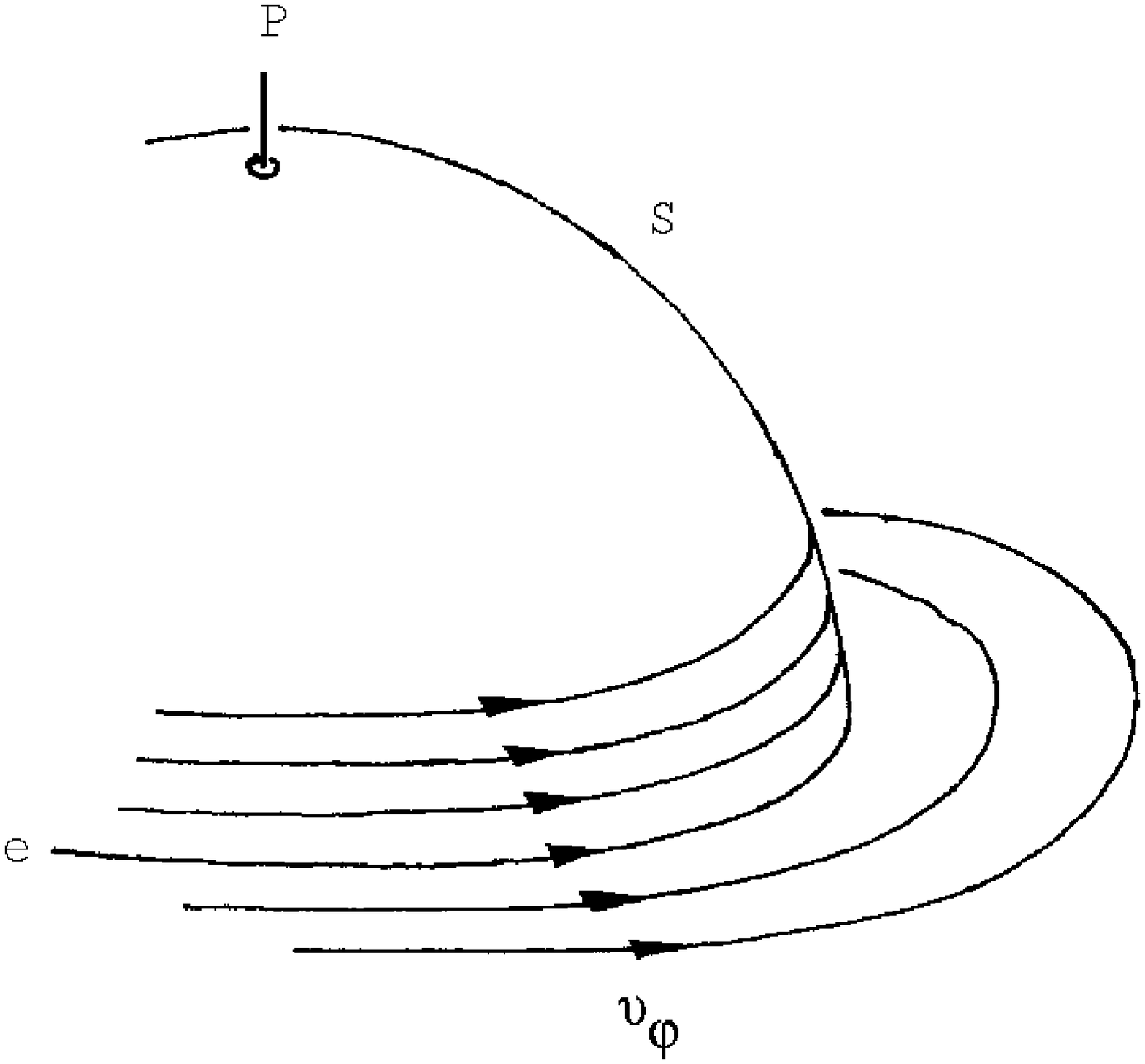}
\caption{Rotation of the matter in the disk
and on the stellar surface
in the proposed model;
$S$ is the stellar surface,
$p$ is the pole,
and $e$ is the equator.}
\end{figure}

In this paper,
we take a completely different approach to the problem.
The deceleration of rotation is considered simultaneously
with the spread of accreted matter (see Figs. 2-4).
Discarding popular beliefs,
we assume that the deceleration of rotation
due to the gripping on
(hooking, catching on)
 the slowly moving dense matter
beneath the underlying surface, $S,$
is the principal friction mechanism.
This gripping is due to the turbulent viscosity.
This problem has much in common with the well-studied
(both theoretically and experimentally)
problem of the deceleration of a subsonic or supersonic flow
above a solid surface
{\it (see Subsec. 2.5 for a detailed discussion
of the turbulence description
accepted in this paper).}

\begin{figure}[tbh]
\epsfxsize=16cm
\epsffile{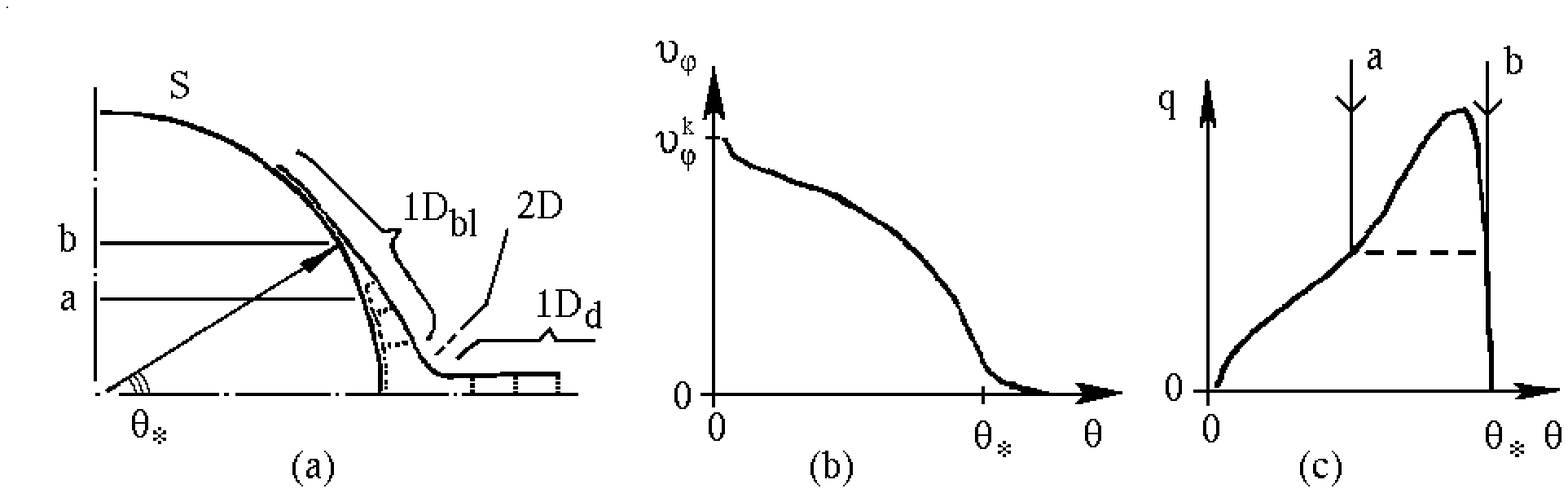}
\caption{Proposed model:
(a) the star is the sphere $S,$
 the one-dimensional spread layer is $1D_{bl},$
 the two-dimensional transition between the disk and the spread layer is $2D,$
 and the one-dimensional disk is $1D_d;$
(b) the linear rotation velocity $v_{\varphi}$ versus latitude.
 The rotation is "localized" in the latitude belt
$0 < \theta < \theta_{\star}$ --
for a nonrotating star,
 the matter outside this belt is essentially nonrotating;
 (c) local radiative flux $q$ versus latitude.
 The arrows $a$ and $b$ "fence" the ring belt of enhanced brightness.
 In Fig. 3a, this belt lies between the latitudes $a$ and $b.$}
\end{figure}

\begin{figure}[tbh]
\epsfxsize=10cm
\epsfbox[-300 197 400 650]{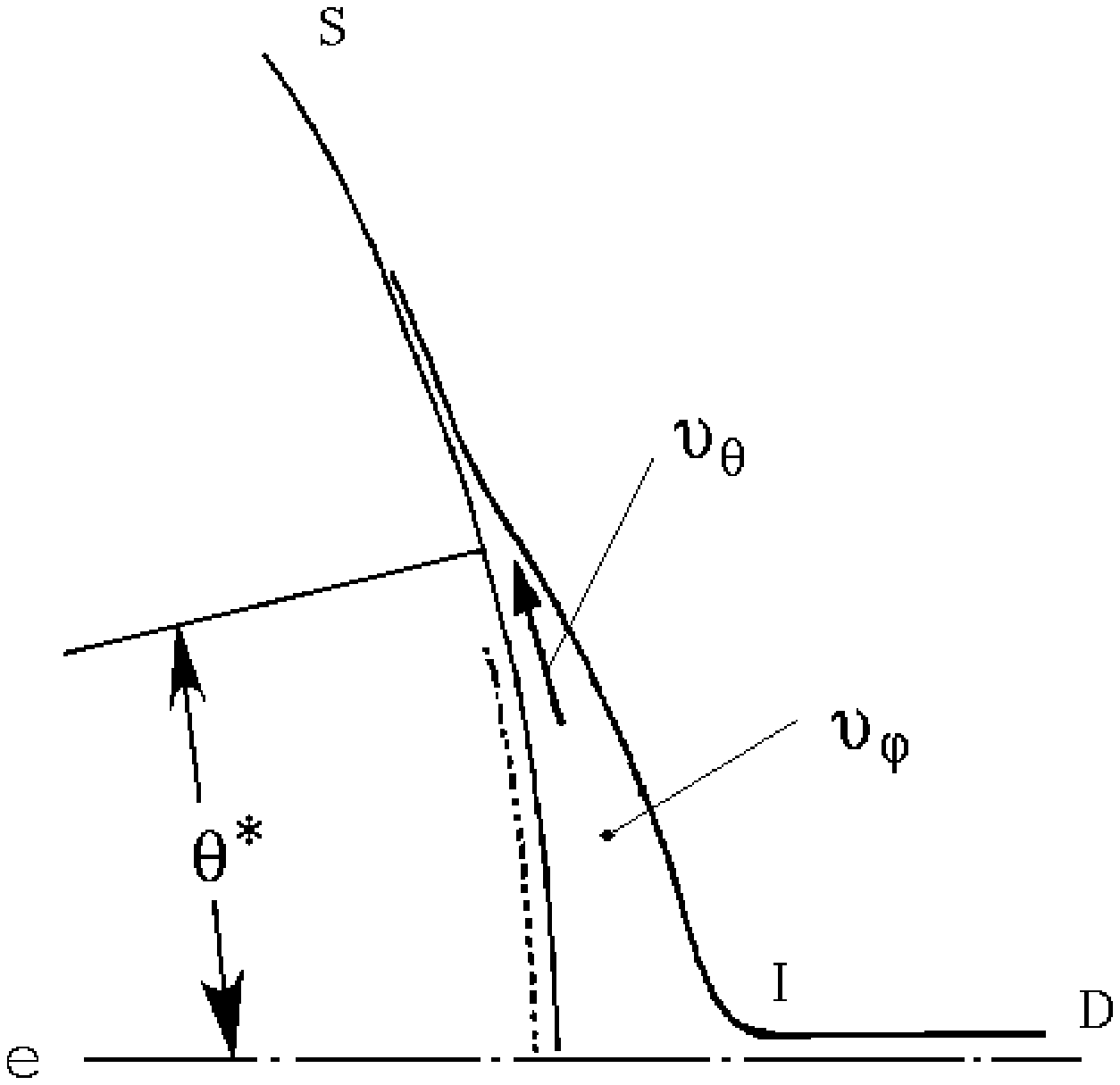}
\caption{Spread of the rotating plasma from the disk, $D,$
 over the neutron-star surface, $S.$
 Here, $I$ is the intermediate zone near the disk neck,
$0<\theta<\theta^{\star}$ is the hot belt,
 and
$\theta > \theta^{\star}$
 is the cold part of the spread layer.
 The rotation velocity $v_{\varphi}$ (filled circle) is directed along the normal to the plane
 of the figure.
 The slowly circulating (in $\varphi$ and $\theta)$
 dense underlying layers of matter beneath the spread layer
 are indicated by the dashes.}
\end{figure}

Here, we ignore the magnetic field, $H,$ of the neutron star.
As we show in Subsec. 4.8,
this assumption is valid for
$H < 2.4\times 10^9 (L_{SL}/L_{edd})^{0.57}$ gauss,
where $L_{SL}$ and $L_{edd}$ are
the spread-layer luminosity and the Eddington critical luminosity,
respectively.
In addition, when deriving the equations of motion of the accreting flow
over the stellar surface,
we disregard its rotation and general-relativity effects.
An appreciable fraction of the X-ray bursters are neutron stars
with a rotation frequency
$f_{ns} = \omega_{ns}/2\pi$
of the order of 300 Hz
see, e.g., a review by Van der Klis 1998).
At the same time, the Keplerian rotation frequency
$f_k = \omega_k/2\pi$
at the disk-star boundary in the problem under consideration
is about 2000 Hz.
Thus,
the effects of stellar rotation,
which are of the order of the ratio $(f_{ns}/f_k)^2,$
are small,
and the computations
which are performed below
also yield an acceptable first approximation
to the problem of accretion onto a rotating neutron star.

\subsection{Comparison of the Standard and Proposed Approaches}

The most important points (I-IV)
that pertain to our comparison of the approaches
are listed below

(I) Let's consider the relationship between the radial
(inflow to the star, $v_r < 0)$
and meridional $(v_{\theta})$
velocities.
The meridional velocity is commonly ignored $(|v_r| \gg v_{\theta}).$
The one-dimensional $(1D)$ approach to the accreting disk
also extends to the boundary layer.
In any case, $v_{\theta}$ is generally omitted from the equations.
The boundary layer turns out to be an extension of the disk
in the sense that the $1D$ description of the disk $(1D_d)$
also applies to the boundary layer.
The disk thickness $h$ passes through a minimum near the neck (see Fig. 1a).
After the neck, the flow broadens
as it approaches the stellar surface
and 'digs itself' into the star.

Meanwhile, a flow with an approximately $1D$ disk,
with a $2D$ region of a sharp turn of the flow near the neck,
and with an approximately $1D$ (hydraulic) spread of the boundary layer
over the neutron-star surface 
seems more natural (see Figs. 1, 3, 4 and Subsec. 2.1).
In this case, $v_{\theta} \gg |v_r|$ in the boundary layer.
We specify inflow conditions
at the interface between the boundary layer and the $2D$ region
when calculating the former,
because we do not consider here an exact $2D$ problem
of the flow in the intermediate zone.
In this formulation,
it would be more correct to talk not about the {\it boundary layer of the disk}
but about the {\it spread layer.}
The dynamics of the latter bears no relation to the disk.
Here, the problem of the deceleration and spread
of the rapidly rotating matter
arises.

(II) We consider the deceleration of rotation
and the structure of isolines of rotation velocity $v_{\varphi}.$
In the standard and proposed approaches,
we have
$v_{\varphi}=v_{\varphi}(r)$
and $v_{\varphi}=v_{\varphi}(\theta)$ in the boundary layer, respectively.
The lines of constant $v_{\varphi}$ (isolines)
are perpendicular to the equatorial plane in the standard model
and perpendicular to the surface $S$ in the proposed model.
In both cases, the rotation is Keplerian in the disk at $r > r_n,$
where $r_n$ is the neck radius.

The $v_{\varphi}$ isolines are indicated in Fig. 3a by the dots.
They are roughly perpendicular to the surface $S$ inside the boundary layer.
The isolines make a sharp turn near $S$
and then run in a dense bundle beside $S$ for a star at rest.
This implies that the rotation velocity 
$v_{\varphi} = \omega r$ abruptly decreases
near the interface between the spread layer and the underlying surface.
The deceleration near the surface $S$ is discussed in Subsec. 2.5.

(III) In the standard models,
the boundary layer and main energy release
concentrate near the equator (see Fig. 1).
In the proposed model,
the width $\theta_{\star}$ of the rotating belt is variable.
It increases with increasing accretion rate, $\dot M.$
As $\dot M \to \dot M_{pole},$
the rotation extends to the entire stellar surface:
$\theta_{\star}\to 90^0.$
Calculations show that the spread-layer luminosity
which corresponds to this accretion rate
is 
$L_{pole} \approx 0.9 L_{edd,}$
where
$$
L_{edd}= 4\pi R^2 q_0 = 1.26 \times 10^{38} \frac{M}{M_{\odot}},
\; \mbox{erg/s},
$$
$$
q_0=
\frac{m_p \, g_0 \, c}{\sigma_T}=
\Sigma_T \, g_0 \, c =
1.00\times 10^{25}\frac{(M/M_{\odot})}{R_6^2},
\; \mbox{erg/s cm}^2,
\eqno (1.1)
$$
where $g_0 =\gamma M/R^2 =(v^k_{\varphi})^2/R,$
$v^k_{\varphi}$ is the Keplerian velocity at the stellar surface,
$\Sigma_T =m_p/\sigma_T = 2.5$ g/cm$^2$
defines a column of matter with a unit optical depth for Thomson scattering,
$m_p$ is the proton mass (the plasma is assumed to be hydrogenic),
$\sigma_T = (8\pi/3)(e^2/m_e c^2)^2$
is the Thomson scattering cross section,
and $R_6 = R/(10$ km).
Below, we talk only about the spread-layer luminosity
at the stellar surface normalized to the critical Eddington luminosity.
Clearly, the accretion disk makes a comparable contribution
to the system's total luminosity.

(IV) We consider the latitude dependence of the intensity of radiation $q$
from the belt.
In the standart approach,
the function $q(\theta)$ has a maximum on the equator.
In the approach under discussion,
$q(\theta)$ turns out to have a minimum at $\theta = 0^0.$
Moreover, $q(\theta = 0) \approx 0.$
The distribution of $q$ (see Fig. 3c) resembles the letter M,
if the plot is complemented with the half
which is symmetric around the equator.

\subsection{Characteristic Features of the Model}

The problem of the spread of rapidly rotating accreting matter
over a neutron-star surface
has several important features.

(I) The centrifugal force
which acts on the accreting matter in the vicinity of the equator
essentially offsets the gravitation,
because the accreting matter in this vicinity
has a velocity that is approximately equal to the Keplerian velocity.
As one recedes toward the pole,
the centrifugal force decreases due to the deceleration of rotation.

(II) In the radiating belt,
the radiation pressure works against the difference
between the gravitational component
and the normal component of the centrifugal force.
Because of the rapid rotation,
the local Eddington luminosity
at which the radiation-pressure force equals the pressing force
is appreciably lower than the luminosity $q_0$
calculated from the standard formula (1.1).

(III) The dissipation of rotational kinetic energy
causes a strong energy release near the lower boundary of the spread layer.
As a result, powerful radiation,
which diffuses through the spreading plasma
and forms the observed spectrum, is produced in this sublayer.
The normal component of the gradient in the radiation pressure
(or the normal component of the radiation-pressure force)
is opposite the direction of the gravitational force.
Its direction coincides with the direction
of the normal component of the centrifugal force.
Within the layer,
the normal component of the radiation-pressure gradient
(together with the centrifugal force)
decreases the free fall acceleration,
decreases the density of the matter in the layer,
and increases its thickness.
An important point is that the increase in the height of the column
(in the spread-layer thickness)
and the drop in the plasma density
cause the efficiency of the turbulent friction
of the layer against the stellar surface
to decrease.
As a result, the deceleration of Keplerian rotation
occurs on larger meridional scales
than that for a small radiation-pressure force.

(IV) The numerical solution of the equations of radiation hydrodynamics,
which is presented below,
gives the following picture.
The local radiative flux, $q,$ at the equator
is much lower than the Eddington critical value (1.1).
This is attributable to the action of the centrifugal force
and to the fact that the rotation velocity $v_{\varphi}$
near the equator is close to the Keplerian velocity.
A ring zone
in which the energy release increases
due to the decrease in the centrifugal force
is formed at a distance from the equator along the meridian.
Up to $70\%$ of the layer luminosity is released
in two such latitude rings (see Figs. 3a and 3c).
In this case, the local flux $q$
never reaches the Eddington limit (1.1),
because the finite rotation velocities $v_{\varphi}$
are required for the energy release
through friction against the underlying layers,
while the allowance for the centrifugal force turns out to be important.
The flux $q$ at maximum depends on accretion rate
and accounts for $70-97\%$ of the limit (1.1).

(V) The spreading plasma within the bright belt is radiation-dominated
$(p_r \gg p_{pl}).$
The high saturation of the radiating belt with radiation
causes the speed of sound, $c_s,$ in the layer,
which proves to be much larger
than the speed of sound in plasma without radiation,
to increase significantly.
As a result,
the flow rotation velocity, $v_{\varphi},$ in the layer
is moderately supersonic:
$\Ma_{\varphi} = v_{\varphi}/c_s \sim 5.$
The meridional spread is subsonic,
which is important in choosing a friction model.
If the Mach numbers, $\Ma_{\varphi},$ were calculated
from the plasma speed of sound alone,
then the flux would be hypersonic with $\Ma_{\varphi} \sim 100;$
moreover, the flux in $v_{\theta}$ would be transonic.

(VI) At $\theta > \theta_{\star},$
the velocity, $v_{\varphi},$ drops virtually to zero,
and the layer contracts to a thin, cold weakly radiating film
('dark' layer),
which slowly moves toward the poles.
Having lost the excess (relative to the star) angular velocity
in the radiating belt,
the matter becomes cold and dense.
The intensity of the radiation from its surface
decreases by several orders of magnitude.
It moves relative to the star with a velocity that is considerably lower
than the speed of sound.
Of course, this matter is not accumulated near the poles.
There is a slow flow
in which it spreads under the hot layers over the entire stellar surface.
This matter forms the surface
against which the radiation-dominated layer rubs.

(VII) If the local radiative flux were the limiting Eddington flux (1.1),
the width of the radiating belt
(let us denote this width by $\theta^{\star}_0$
to distinguish it from $\theta_{\star})$
could be estimated from the energy balance
$$
(1/2)\,\dot M \,( v_{\varphi}^k )^2 =
q_0 \, 4\pi R^2 \,\sin \theta^{\star}_0,\;\;\;\;\;
\theta_0^{\star} = \arcsin ( L_{SL}/L_{edd} ), \eqno (1.1)'
$$
where $L_{SL}$ is the spread-layer luminosity.
As will be seen from the analysis in the main body of this study,
$(1.1)'$ underestimates appreciably $\theta_{\star}.$
This is attributable to the action
of the radial component of the centrifugal acceleration.
As follows from the calculations which are given below,
the approximate balance is
$$
\dot M\,\frac{(v_{\varphi}^k)^2}{4} =
2 \pi R^2 \, \int_0^{ \theta_{\star} }
\, q_{eff} (\theta) \; d\sin \theta, \eqno (1.2)
$$
$$
q_{eff} (\theta) = \Sigma_T \, g_{eff}(\theta) \, c, \;\;\;
g_{eff}(\theta) = g_0 \, G_{eff} (\theta),\;\;\;
G_{eff} = 1 - U^2 - W^2 \approx 1-W^2, \eqno (1.2)'
$$
where $U = v_{\theta}/v_{\varphi}^k,$
$W = v_{\varphi}/v_{\varphi}^k,$
$v_{\varphi}^k$ is the Keplerian velocity.
In $(1.2)',$ we assumed
that the centrifugal force
is almost completely determined by the rotation velocity, $v_{\varphi},$
because the centrifugal force related to the spread velocity is small:
$v_{\theta}^2/R \ll g_0.$
The latitude dependence of rotation velocity
$v_{\varphi}( \theta )$ is required to calculate $\theta_{\star}$ using (1.2).
The approximate validity of relations (1.2) and $(1.2)'$
is attributable to the fact
that the Eddington radiative flux $q_r$ in the radiating belt
offsets the difference $g_{eff}$
between the weight and the radial component of the centrifugal force
with a high precision [see below Subsec. 1.3 (I)].
The appearance of rings of enhanced luminosity follows precisely from this.
Indeed, it follows from $(1.2)'$
that $q( \theta )$ increases with $\theta,$
because $v_{\varphi}( \theta )$ decreases with latitude.

(VIII) We carried out this study
in an effort to explain the observed spectra of bursters
in a wide range of their luminosities.
Several variable sources that are X-ray bursters exhibit variations
in the X-ray luminosity by hundreds of times
(see Campana et al. 1998a, 1998b, Gilfanov et al. 1998).
The theory,
 which we construct here,
claims to describe some of the spectral features in the light curve
of a neutron star as the luminosity changes by two orders of magnitude.

(IX) The atmosphere under consideration
is characterized by the delicate balance of three forces:
gravitation, centrifugal force, and radiation-pressure force.
Under these conditions
(just as in the atmospheres of massive hot stars),
a strong wind
which flows away from the zone of enhanced brightness
is inevitably formed.
As a first approximation,
we assume below
that this wind affects only slightly the transport characteristics
of the spread layer and its optical depth.

(X) It follows from the qualitative estimates
and the detailed calculations
which are presented below
(see Subsecs. 4.3 and 4.5)
that the derived solutions in the radiating belt
lie near the limit of validity of the shallow-water approximation.
The ratios of the effective layer thickness,
$h_{eff} = [\,|\, d \ln \rho/dr \,|\, ]^{-1},$
in this region
(the derivative is calculated near the base of the spread layer)
to the stellar radius, $R,$ are typically $\sim 0.1.$

\begin{figure}[tbh]
\epsfxsize=10cm
\epsffile{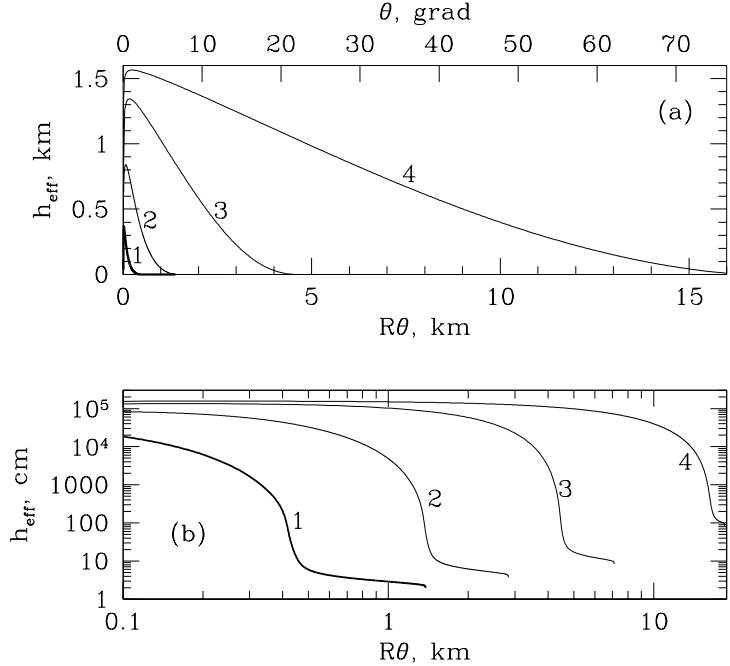}
\caption{Profiles 1, 2, 3 and 4 of the spread-layer (S.L.) effective thickness
$h_{eff}(\theta)$
 for various accretion rates $\dot M.$
 The relative luminosities, $L_{SL}/L_{edd},$
 that correspond to labels 1, 2, 3 and 4 are 0.01, 0.04, 0.2 and 0.8, respectively.
 The stellar mass and radius are
$1.4 M_{\odot}$
 and 12 km.
 (a) Linear scales, the shape of the radiating belt is seen;
 (b) logarithmic scales, the dark layer  and the spread layer
are seen better 
than in the upper panel
 (especially at small $\dot M;$
the scale $R \theta$ covers the entire distance from the equator to the pole).}
\end{figure}

Nevertheless, the hydraulic approximation
(Schlichting 1965; Chugaev 1982)
describes satisfactorily the experimental situation
at the values of $h_{eff}/R.$
The geometrical characteristics of the spread layer are shown in Fig. 5.
The spread layer is divided into two distinctly different parts
at $\theta_{\star}:$
the radiating belt (r.b.) and the dark layer (d.l.).
Their thicknesses, $h_{rb}$ and $h_{dl},$ differ markedly.
The thickness $h_{dl}$
is of the order of the scale height
in a gravity field
with $g_0=\gamma M/R^2.$
The slope $\tan \beta$ is $dh_{eff}/R d\theta.$
This slope changes only slightly inside the radiating belt (see Fig. 5a).
The angle $\beta$ is very small in the dark layer.
The angle $\beta$ in the radiating belt
increases as the accretion rate, $\dot M,$ decreases.
Therefore flows with low $\dot M$ are described worse
by the $1D$ approximation.
In Figs. 5a and 5b, profiles 1-4 refer to the different values of $\dot M.$
Profiles 1-4 from Fig. 5a are repeated in Fig. 5b on a logarithmic scale.
This is necessary
because the dark-layer thickness, $h_{dl},$
is so small
that it cannot be compared with the radiating-belt thickness $h_{rb}$
on a linear scale.
In addition, the flow separation into two characteristic regions
is more clearly seen in Fig. 5b.
The latitude $\theta_{\star}$
corresponds to the thin transition zone of a rapid change in thickness.
In turn, the linear coordinates in Fig. 5a are required
to show the shape of the radiating belt.

(XI) The flows in the disk and in the spread layer are closely related.
The accreting flow is furnished through the disk,
while irradiation of the disk from the stellar surface
affects its characteristics.
This issue is analyzed in Subsec. 2.1,
where the disk thickness in the zone of its contact
with the spread layer is estimated.

(XII) In a radiation-dominated layer,
the hydrodynamic meridional transfer of radiative enthalpy toward the poles
plays a major role.
This enthalpy per unit mass
is of the order of the gravitational energy $\gamma M/R.$

\subsection{Main Results}

We derive the system of equations of radiation hydrodynamics
that describes the deceleration of rotation of the accreting matter
and its meridional spread.
The derivation is based on the assumption
that Thomson scattering contributes mainly to the opacity
and that the energy release concentrates in a relatively thin sublayer
at the base of the spread flow.
The derived equations {\it differ markedly}
from the standard shallow-water equations.
A qualitative analysis of the derived system of equations
and their solutions (Sec. 3) allows us,
first, to develop a procedure for the effective numerical integration
of the equations
and,
second,
to make a clear interpretation of the calculations.
We carried out a numerical analysis of the system.
The problems of the numerical analysis,
without which an independent verification of the data would be impossible,
are outlined in Secs. 3 and 4.
It is important to set out the procedure for the solution,
because if this procedure is clear
the results can be reproduced.
We thus would like to draw the reader's attention
to the following subsections:
(I) smallness of the coefficients,
rigidity of the system,
steep and gentle segments (Subsec. 3.5);
(II) multiplicity of the solutions,
selection criteria (Subsecs. 3.6 and 4.1-4.4);
(III) circulation of the cooled accreting matter
which lost the rotational velocity component,
its spread and settling (Subsecs. 4.2-4.4);
and (IV) modification of the solutions
as the friction coefficient and the accretion rate are varied
(Subsecs. 4.5 and 4.7).

Below, we dwell on the most important results.

(I) {\bf Rings of enhanced brightness.}
Typical latitude profiles of the flux $q$ are shown in Figs. 6-8.
The radiating belt broadens as the accretion rate increases.
The latitude $\theta_{max}$
at which the bright rings lie
increases with increasing $\dot M$ and $L_{SL}/L_{edd}.$
The bright rings are not narrow;
their width is $\Delta \theta \simeq \theta_{\star}.$
At a high accretion rate $\dot M \simeq \dot M_{pole}$
[see Subsec. 1.1 (III)],
the entire stellar surface,
except the small equatorial and polar zones,
emits almost uniformly.
In this case, the bright ring spreads out.
Profile 4 in Fig. 7 refers to this situation.

\begin{figure}[h]
\epsfxsize=7.cm
\epsfbox[-250 180 430 630]{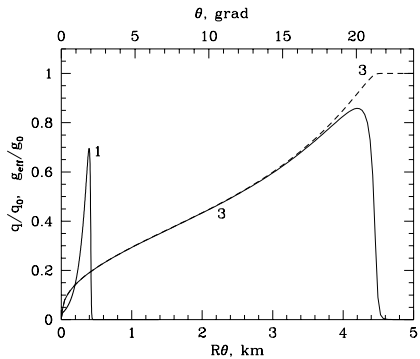}
\caption{Local radiative flux $q(\theta)$ (solid lines)
 and effective free fall acceleration (including the centrifugal acceleration) 
$g_{eff} (\theta)$
 (dashed lines) versus latitude.
 An important point is that
$q/q_0\approx g_{eff}/g_0$
 in the radiating belt [see (1.2) and (1.2)'].
 Labels 1 and 3 refer to
$L_{SL}/L_{edd} =  0.01$ and $0.2,$
respectively.}
\end{figure}

\begin{figure}[h]
\epsfxsize=7.5cm
\epsfbox[-250 180 430 590]{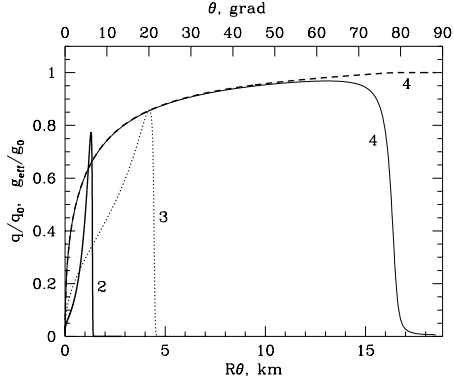}
\caption{Changes in the profiles of the local radiative flux $q$
 (solid lines)
 with $\dot M.$
 The radiating belt broadens with increasing $\dot M.$
 The boundary 
$\theta_{\star}$
 lies at the right edge of the belt in the zone of an abrupt decrease
 in $q(\theta).$
 The $g_{eff}/g_0$  profile at $L_{SL}/L_{edd}=0.8$
 is indicated by the dashed line.
 The scale $R \theta$ terminates at the pole.
 Labels 2, 3, and 4 refer to
$L_{SL}/L_{edd} = 0.04, 0.2,$ and 0.8, respectively.}
\end{figure}

\begin{figure}[tbh]
\epsfxsize=7.2cm
\epsfbox[-250 180 430 665]{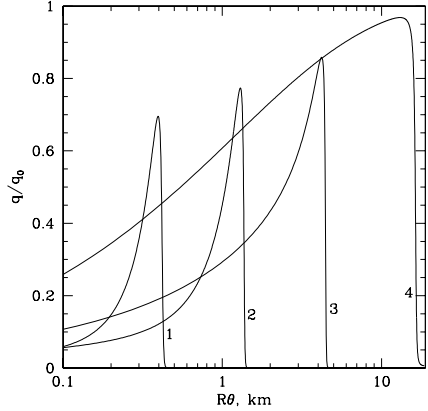}
\caption{$q( \theta )$ profiles in a wide range of $\dot M.$
Labels 1, 2, 3, and 4
refer to $L_{SL}/L_{edd} = 0.01, 0.04, 0.2,$ and $0.8,$
respectively.}
\end{figure}

The function $q(\theta),$
which gives the radiative flux,
has the maximum at $\theta_{max}$
with which the ring of enhanced brightness is associated.
This ring is marked by arrows $a$ and $b$ in Fig. 3c.
The ring bounded by latitudes $a$ and $b$
which correspond to these arrows
is shown in Fig. 3a.
To the right of $\theta_{max},$
the rotation velocity $v_{\varphi}$ vanishes (see Figs. 3b and 3c).
For this reason,
$q( \theta )$ rapidly decreases to the right of $\theta_{max}.$
The increase in $q( \theta )$ to the left of $\theta_{max}$
(see Fig. 3c)
with increasing $\theta$ is caused by the decrease in $v_{\varphi}$
(see Fig. 3b).
Indeed, the decrease in $v^2_{\varphi}$
leads to an increase in the acceleration
$g_{eff} \approx g_0 \, (1-W^2),$
where $W=v_{\varphi}/v_{\varphi}^k,$
and, consequently, to an increase in $q,$
$\; q \propto g_{eff}$ [see formulas (1.2) and $(1.2)'].$

\begin{figure}[tbh]
\epsfxsize=8cm
\epsfbox[-250 180 360 650]{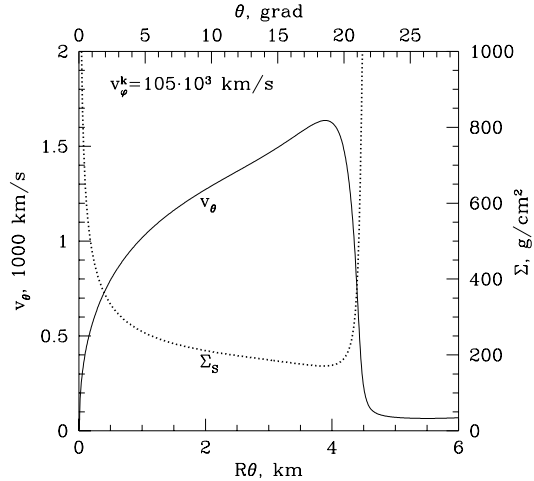}
\caption{Profiles of the meridional velocity $v_{\theta}$
 and of the spread-layer surface density $\Sigma_S$ for
$L_{SL}/L_{edd}=0.2.$
 At $\theta_{\star},$
 the flow abruptly slows down,
 and its density increases.
 In the upper corner, the Keplerian velocity
$v_{\varphi}^k$ is shown for
$M = 1.4 M_{\odot}$
and $R = 12$ km.}
\end{figure}

(II) {\bf Surface density, optical depth, and spectral hardness.}
In the steady-state case
in the absence of sources and sinks of matter in the spread layer,
the flux of mass is conserved
when spread along the meridian.
We ignore the settling of matter to the bottom within this layer
(Subsecs. 4.3 and 4.4).
The meridional velocity is thus $v_{\theta} \propto 1/\Sigma_S,$
where $\Sigma_S = \int \rho dr$ is the surface density,
the integral is taken over the spread-layer thickness.
Accordingly, $\Sigma_S$ reaches a minimum at the maximum of $v_{\theta}.$
These profiles are illustrated in Fig. 9.
The spread layer consists of the radiating belt and the dark layer
which border at $\theta_{\star}.$
The maximum of $v_{\theta}$ and the minimum of $\Sigma_S$
lie near the edge of the radiating belt.
In the dark layer, $\Sigma_S$ abruptly increases,
while $v_{\theta}$ abruptly decreases.
This is clearly seen in Figs. 9 and 10.
The dark layer terminates at the sonic point.
This important problem is analyzed in Subsecs. 4.1-4.4.

Figure 10 shows
how the spread-layer surface density $\Sigma_S$ depends on accretion rate.
The spread-layer optical depth $\tau_T = \Sigma_S/\Sigma_T$
for Thomson scattering
rapidly decreases with decreasing $\dot M.$
At $L_{SL}/L_{edd}\sim 10^{-2},$ we have $\tau_T \sim 3.$
No blackbody radiation can be produced
in the medium at such a small optical depth;
the protons and electrons have temperatures much higher than 1 keV.
The low-frequency photons emitted by the dense underlying layers
are comptonized by hot electrons of the spread layer
and form the experimentally observed hard power-law tails in the spectra
(Barret et al.1992, Campana et al. 1998a, 1998b, Gilfanov et al. 1998).
In turn, some of the hard photons penetrate deep into the underlying layer
and heat it up
[see Sunyaev and Titarchuk (1980) for a discussion of this problem].

\begin{figure}[tbh]
\epsfxsize=8cm
\epsfbox[-220 180 400 630]{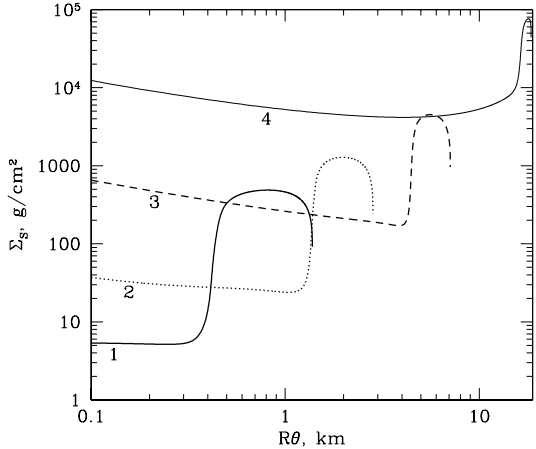}
\caption{Shape of the $\Sigma_S( \theta )$ profile
at different spread-layer luminosities.
At a fixed $\dot M,$
the surface density rises with increasing $\theta$
at $\theta_{\star},$
after which an optically thick weakly radiating layer begins.
Labels 1, 2, 3, and 4
refer to $L_{SL}/L_{edd} = 0.01, 0.04, 0.2,$ and $0.8,$ respectively.}
\end{figure}

At a high spread-layer luminosity $L_{SL} > 0.1 L_{edd},$
we have a different picture.
In this case, the spectrum is much softer and exhibits no pronounced hard tail.
The optical depth for Thomson scattering is great $(\tau_T \sim 1000).$
The layer optical depth for effective absorption
$\sqrt{\tau_T \; \tau_{ff} (\nu, T_e)}$
(Shakura 1972; Felten and Rees 1972; Shakura and Sunyaev 1973;
Illarionov and Sunyaev 1972)
turns out to be also large.
Deep in the layer,
the most important physical process is saturated comptonization.
The parameter $y = (T_e/m_e c^2) \, \tau_T^2,$
which characterizes comptonization in a layer with an optical depth $\tau_T,$
is large even at $\tau_T > \sqrt{m_e c^2/T_e} \sim 20.$
At $T_e \sim T \sim \tau_T^{1/4}$ (see Subsec. 2.2)
and $\tau_T \sim 10^3,$
we have $y \sim (T_S/m_e \, c^2)\, \tau_T^{9/4} \sim 10^5$ deep in the layer.
The dimensionless frequency,
near which the rate of photon absorption due to the free-free processes
equals the rate of photon removal upward along the frequency axis
via comptonization,
is
$x_0 = h \nu_0/T_e =3 \times 10^5\, n_e^{1/2}/T_e^{9/4}.$
This frequency,
deep in the layer,
is high.
Here, the electron density is in cm$^{ - 3},$
and the electron temperature is in Kelvins.
As can be seen from the plots in Illarionov and Sunyaev (1975),
comptonization produces the Wien spectrum deep in the layer at $y>10,$
while at $y > 10^5$ the spectrum is similar to a blackbody
due to the comptonization of low-frequency photons.
Consequently,
the combined effect of bremsstrahlung and comptonization
produces a blackbody spectrum deep in the layer.
The emission from the dense underlying layers
is yet another source of soft photons for comptonization.
Nevertheless, the emergent spectrum differs markedly from a blackbody spectrum,
because at each frequency we see a portion of the spectrum
that is produced at its own depth
[see Illarionov and Sunyaev (1972) for a discussion of this problem].

{\bf Integrated spectrum of the spread layer.}
The spread-layer spectrum is given by the integral
of the emergent spectra at each point of the layer over its surface.
Here, it is important
that the spectrum and the local radiative flux depend on latitude.
The integrated spectrum is not presented
for the following two reasons.

(a) A considerable fraction of the spread-layer radiation
falls on the disk surface (see Fig. 2).
This fraction is partially reflected from the disk
and partially reprocessed by it into softer radiation.
The intensity of the disk-reprocessed radiation
depends on the binary's inclination (see Lapidus and Sunyaev 1985).
Allowance for these effects requires a special study.

(b) Because of the shielding by the disk,
the observer cannot see the lower half of the neutron star.
Part of its surface is located in the shadow region
and does not contribute to the direct rays viewed by the observer
(here, we disregard the general-relativity effects).
In Fig. 11,
the flux of "direct" photons from the stellar surface is plotted
against the cosine of the binary's inclination angle, $i,$
which is measured from the star's polar axis (see Figs. 1-4).
In these calculations,
we assumed the indicatrix of emission per unit area to be
$\varphi (\mu) = 1 + 2.06 \, \mu
\;(\mu = \cos i);$
this is the first approximation for the angular dependence
of the intensity of the emergent radiation from a scattering atmosphere
(Sobolev 1949; Chandrasekhar 1960).
In the calculations, we took into account the profiles $q( \theta )$
(see Figs. 6-8).

\begin{figure}[tbh]
\epsfxsize=9cm
\epsfbox[-220 170 392 700]{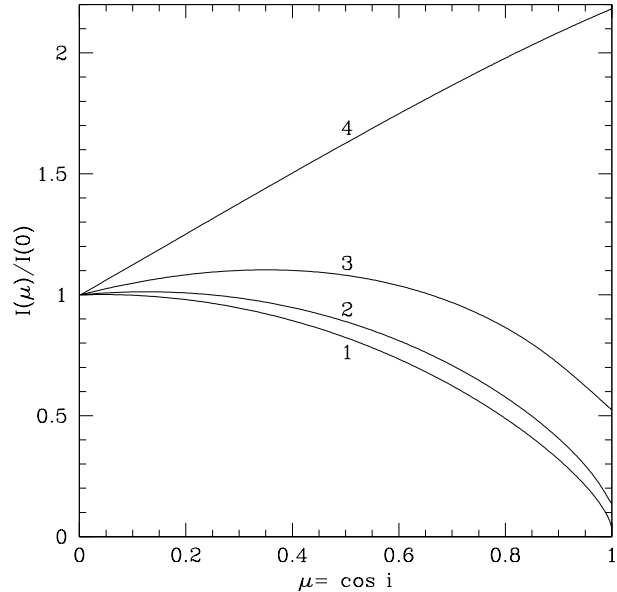}
\caption{Functions $I(\mu)/I(0)$
$ [I$(erg s$^{ - 1}$ steradian$^{ - 1},$ $\;\mu = \cos i]$
 which give the intensity of the radiation from the visible neutron-star surface.
 The binary's inclination, $i,$ is measured from the polar axis.
 Labels 1, 2, 3, and 4 refer to
$L_{SL}/L_{edd} = 0.01, 0.04, 0.2,$ and $0.8,$ respectively.
 The functions are normalized to the intensity of the radiation from the equator $I(0).$}
\end{figure}

Clearly, the presence of a "shadow" and the reflection by the disk
must lead to an appreciable polarization of the emission
from an accreting neutron star.
At low luminosities,
at which the radiating belt is bound by the equatorial zone,
the flux toward an observer in the equatorial plane $(\mu = 0)$
exceeds the flux toward the polar axis $(\mu = 1).$
Curves 1-3 in Fig. 11 correspond to these luminosities.
At high luminosities, the emission extends to the entire stellar surface
(curve 4 in Fig. 7).
In this case,
the flux toward the polar axis is greater
than the flux toward the equatorial plane
(curve 4 in Fig. 11).
Indeed, the ratio $I(1)/I(0) = 2$ for a uniform emission
of the stellar surface,
because the disk occults the lower stellar hemisphere.

(III) {\bf Deceleration efficiency.}
The meridional distributions of rotation velocity $v_{\varphi}$
are shown in Fig. 12.
We performed the calculations for $\delta = 10^{-2},$
where $\delta = 1 - W_0 = (v^k_{\varphi} - v^0_{\varphi})/v^k_{\varphi},$
where $v^0_{\varphi}$ is the initial rotation velocity in the spread layer
when passing from the disk to the layer,
and $v^k_{\varphi}$ is the Keplerian velocity (see Subsecs. 3.4 and 4.6).
Similar results are also obtained for different values of $\delta \ll 1.$
As $\dot M$ increases,
the deceleration of rotation requires an increasingly large area
of contact
between the radiating belt and the stellar surface.
The boundary of this belt is $\theta_{\star}.$

\begin{figure}[tbh]
\epsfxsize=11cm
\epsfbox[-100 180 450 665]{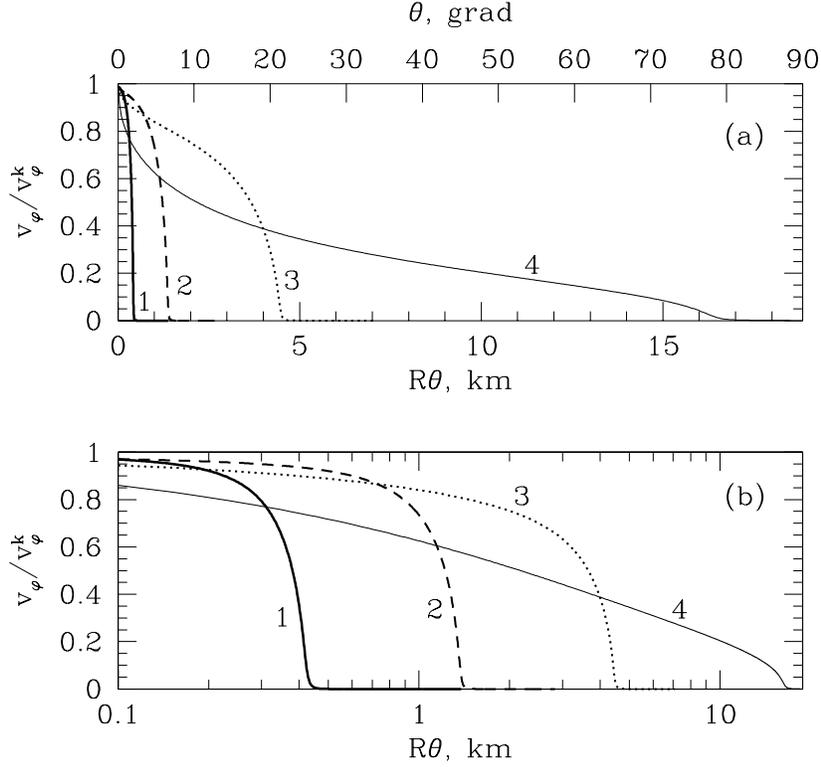}
\caption{Meridional profile of the rotation velocity.
 The radiating belt is simultaneously the rotating belt (cf. Figs. 6-8).
 Labels 1, 2, 3, and 4 refer to
$L_{SL}/L_{edd} = 0.01, 0.04, 0.2,$ and $0.8,$ respectively.}
\end{figure}

An interesting characteristic of the deceleration efficiency
is the number of rotations
a test particle makes around the star
before it looses its rotation velocity
and reaches the latitude $\theta_{\star}.$

\vspace{.2cm}
\centerline{{Table.} Dependencies of
$\theta_{\star},$
$N_{rot}$
and $q/Q^+$ on $L_{SL}/L_{edd}.$}

\vspace{.2cm}
\hspace{4.3cm}
\begin{tabular}{|c|c|c|c|c|}  \hline
$L_{SL}/L_{edd}  $         & 0.01 & 0.04 & 0.2 & 0.8   \\ \hline
$R\theta_{\star},$      km & 0.43 & 1.36 & 4.5 & 16.4  \\ \hline
$N_{rot}         $         & 0.3  & 1.1  & 5.4 & 36    \\ \hline
$(q/Q^+)_{max}   $         & 1.01 & 1.05 & 1.3 & 3.8   \\ \hline
\end{tabular}

\vspace{.2cm}
\noindent The table shows
how the width of the belt $\theta_{\star}$
and the number of rotations,
$$
N_{rot} = \Delta \varphi/2\pi,\;\;\;\;\;
\Delta\varphi = \int_{\theta_0}^{\pi/2} \,\omega \,dt \approx
\int_0^{\theta_0} \frac{v_{\varphi}}{v_{\theta}}\;
\frac{d \theta}{\cos \theta},
$$
of its rotating plasma
depend on the spread-layer relative luminosity $L_{SL}/L_{edd.}$

As we see from the table, at a low spread-layer luminosity $(0.01 L_{edd}),$
a test particle does not make even one rotation around the star,
rising by 430 m along the meridian and covering 23 km in latitude.
Note that the meridional boundary-layer extent (see Fig. 1)
$R \theta_{bl}= R \sqrt{2 T_{bl}/m_p}/v_{\varphi}^k$
is a mere 70 m in the standard approach (at $T_{bl} = 3$ keV).
At a high luminosity, the deceleration efficiency is much lower:
at $L_{SL} = 0.8 L_{edd},$
test particles make 36 rotations around the star
before they slow down and reach $\theta_{\star}.$

(IV) {\bf Advection of radiative energy.}
The spreading matter in an optically thick layer transports radiative energy
along the meridian from the equatorial region
into the region of the bright rings.
The energy transfer is described by the energy equation
(see Subsecs. 2.7 and 3.2).
Our calculations (see Fig. 13) show that the energy flux $q$
emitted per unit area of the spread layer in the equatorial region
is much smaller than the local energy release
through the turbulent friction in a column
$Q^+ = \int \, \dot \epsilon_t \, dy,$
where the integral is taken over the spread-layer thickness
[Subsec. 2.5, (2.17)].
On the other hand,
the flux $q$ at higher latitudes for larger luminosities exceeds appreciably
$Q^+$
[see the table,
which gives the dependence of maximum $(q/Q^+)_{\max}$ in the spread layer
on $L_{SL}/L_{edd}].$
This is only to be expected,
because the spread-layer optical depth for Thomson scattering
is large at high luminosities.
As a result, the characteristic diffusion time $t_d$ of the photons
from the base of the hot layer
turns out to be of the order of the time of hydrodynamic motion $t_h$
through the radiating belt,
where $t_d\sim (h/c)\,\tau_T$
and $t_h \sim R \theta_{\star}/v_{\theta}.$
The advection of radiative energy favors
the formation of bright latitude rings at the stellar surface.

\begin{figure}[tbh]
\epsfxsize=9cm
\epsfbox[-210 279 400 650]{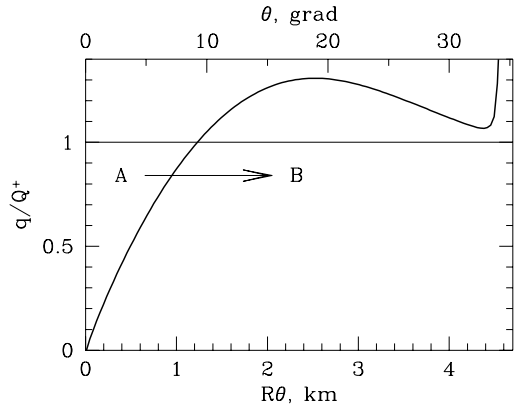}
\caption{Spread dynamics is determined in many ways
by the hydrodynamic transfer of radiative energy.
The ratio of the energy flux, $q,$
emitted per unit area of the radiating layer
to the frictional energy release $Q^+$
per unit area of the contact surface
between the spread layer and the star.
The energy is transferred from zone A into zone B
through meridional advection.
This calculation is for $L_{SL}/L_{edd} = 0.2.$}
\end{figure}

\subsection{Problems That Require Further Studies}

(I) {\bf Possible Instability of the Spread Layer.}
The energy-producing sublayer props up the spread flow from below,
and the radiation pressure dominates.
This may be the cause of convective instability
with floating 'photon bubbles'.
Clearly, this requires that the spread layer be optically thick.
The fluctuations which are produced by the bubbles
may be responsible for the observed low-frequency noise
and even for quasi-periodic oscillations in accreting neutron stars.
Although, on the other hand,
the rotation velocity, $v_{\varphi},$ inside the spread layer
outside the energy-producing sublayer,
along with the normal component of the centrifugal force,
slightly increases with height.
In addition,
the tail\footnote{
The frictional heat release
concentrates mainly in the energy-producing sublayer;
see Subsec. 2.5.}
of the volume density of dissipative energy production
$\dot \epsilon_t$ erg s$^{ - 1}$ cm$^{ - 3}$
extends from the energy-producing sublayer into the spread layer.
As a result,
the flux $q$ and, hence, the Eddington force slightly increase
 with distance from the surface.
These two effects play a stabilizing role
with respect to the convective instability.
The instability is also stabilized
by turbulent viscosity and turbulent diffusion,
which are attributable to the presence of a wall
(the stellar surface from below)
and of shear turbulence.
This turbulence causes, first, the mixing of density fluctuations
(diffusive mixing of photon bubbles)
and, second, the momentum exchange between the bubbles and the ambient medium
(viscous deceleration).
The problem of the vertical stability of hydrostatic quasi-equilibrium,
indubitably,
requires further study.

(II) {\bf Gap between last stable orbit and surface.}
Here, we do not consider the case
where the stellar radius $R$ is smaller
than the radius of the marginally stable Keplerian orbit
$R_c = 3 R_g.$
In this case,
part of the gravitational energy is released in the zone of contact
between the disk and the star
due to an appreciable radial velocity
(it increases with the increasing $R_c - R)$
of the accreting matter on the stellar surface.
The energy release at the equator causes the layer to swell in this zone
because of the radiation pressure,
and the $1D$ approximations appear to become inapplicable.

(III) {\bf Rapid rotation.}
The case of rapid rotation of a star,
where its angular velocity, $\omega_S,$
accounts for an appreciable fraction
of the Keplerian angular velocity, $\omega_k,$
requires a special analysis.

\subsection{Structure of the Paper}

Here, we derive and numerically solve the system of equations of motion
of the matter over a neutron-star surface in the $1D$ approximation.
The vertical (along $r)$
and horizontal (along the meridian $\theta)$ scales
are separated in this approximation.
The gradients in $r$ are assumed to be large
compared to the gradients in the polar angle $\theta.$
For this reason,
the vertical profile is constructed
from the conditions of hydrostatic quasi-equilibrium.
The profile is given by simple analytic expressions.
The averaging of the quantities
over the vertical structure
with which the derivation begins
allows us to obtain a system of equations
for these quantities
that contains only gradients in $\theta.$
These are the equations of radiation-dominated shallow water
that describe the dynamics of the high-velocity layer
in which the newly accreted matter flows over the surface
of the slowly moving high-density underlying layers (Sec. 2).
This completes the part of the paper
in which we formulate the problem.
The part
 in which we present the results
is devoted to the steady-state solutions of our model.
In this part, we describe the physics of adaptation:
effects of the friction model,
Eddington swelling of the boundary layer,
regulation of the deceleration by the swelling, etc. (Secs. 3 and 4).
 We study the mathematical properties of radiation-dominated shallow water:
the structure of the solutions,
the limiting set of steady-state solutions,
its one-parameter structure,
the critical solution, etc. (Sec. 3).
We analyze the dependences of the boundary-layer characteristics
on $\dot M$
and on model parameters (Secs. 3 and 4).
Note that it follows from the solutions
that the bulk of the radiating belt
is roughly in {\it hydrostatic equilibrium along the meridian,}
as was previously suggested by Shakura and Sunyaev (1988).
In this case,
the "pressing" force to the equator is the tangential component
of the centrifugal force,
which counteracts the meridional component of the pressure gradient.

The interaction of the spreading flow with the underlying matter
is described in Subsec. 2.1.
Here, we also estimate the effects of disk heating
by the spread-layer radiation
and determine the minimum possible disk thickness near the neck.
The neck forms a kind of base of the disk
at the location of its interaction with the spread layer.
In Subsec. 2.2, we determine the structure
of a radiation-dominated atmosphere
in which the scattering by free electrons is a major contributor
to the opacity.
The force and energy characteristics of this atmosphere
which are required to calculate its energy and transport properties
are derived in Subsec. 2.3.
A simple formula for radiative losses in the spread layer
is written out in Subsec. 2.4.
Subsec. 2.5 deals with the friction model.
In Subsecs. 2.6 and 2.7, we derive the equations of "radiation" shallow water.
Transformations of this system of equations are made in Subsecs. 3.1-3.3.
In Subsecs. 3.4 and 3.5,
we study the phase space of the system
and introduce an important surface -- the "levitation" surface.
The dependence of the spread velocity on the Mach number
is analyzed in Subsec. 3.6.
The solutions of the derived system of equations
are analytically and numerically analyzed in Subsecs. 4.1 and 4.2.
Subsec. 4.3 investigates the critical regime
in which the spread velocity reaches the speed of sound
for the spread of matter at the edge of the radiating belt.
The dynamics of the matter spread outside the radiating belt
is analyzed in Subsec. 4.4.
The effect of friction on the results is considered in Subsec. 4.5.
Such an analysis is necessitated by the fact that,
first, the friction coefficient is determined only approximately
and, second, this coefficient decreases by a factor of 1.5 to 2
in the supersonic regime (see Subsec. 2.5).
We construct the dependence $q(\theta)$ in Subsec. 4.6
and study how the spread flow is modified
as the accretion rate $\dot M$ is varied in Subsec. 4.7.
At large $\dot M,$
the "squeezing" of the flow due to the convergence of meridians to the pole
becomes important.
In addition, the spread flow in the radiating belt
is strongly subsonic at high $\dot M$
and transonic at low $\dot M.$
In Subsec. 4.8, we estimate the limiting magnetic field strength
that can still be ignored
when the deceleration and spread are studied.
The results are summarized in the conclusion.

\section{HYDRAULIC APPROXIMATION}

\subsection{Spread Scheme}

The spread flow is illustrated in Fig. 4.
Here, $e$ is the equatorial plane,
$D$ is the disk,
$I$ is the transition region
between the disk and the boundary layer
(the region of $2D$ flow),
and $S$ is the stellar surface.
Let $h(\theta)$ be the boundary-layer thickness.
For the $1D$ approximation to be valid, it is necessary that
$$
h\ll R, \;\;\;\;\;\;\; \;\; |dh/d l| \ll 1, \eqno (2.1)
$$
where $l = R \theta$ is the arc length along the meridian.
Note, incidentally, that it follows from the calculations (see below)
that the angle $\arctan |dh/dl|$ is constant
in order of magnitude at $0 < \theta < \theta_{\star}.$

The latitudinally and meridionally moving spread layer
draws (grips) the thin underlying layers into motion
via turbulent friction.
The underlying layer is indicated by the dashed line in Fig. 4.
The $v_{\varphi}$ distribution inside the boundary layer is roughly uniform
in height (see Subsec. 2.5 for a discussion).
Almost all of the drop in $v_{\varphi}$ takes place near the surface, $S,$
which is the bottom of the boundary-layer,
$r = r_S$ or $y = y_S.$
The main turbulent energy release occurs near the bottom.
Since the problem is a steady-state problem,
the temperature $T$ is constant at $y < y_S$
(these layers were heated earlier,
and the energy release in them is small),
while the vertical radiative flux $q$
is approximately constant above $y_S$
and equal to zero below.
Because of the abrupt change in $q$ at the surface $S,$
the plasma parameters also change when this surface is crossed.
The underlying-layer density $\rho_2$ exceeds the boundary-layer density
in the hot subregion
$0 < \theta < \theta_{\star}.$
In the region $\theta > \theta_{\star},$
the density of the underlying layers rapidly increases with depth,
because they are isothermal.
For this reason,
the velocity in the underlying layers rapidly decreases with depth.

{\bf Disk heating by spread-layer radiation.}
Let us find out
how the disk thickness near a neutron-star surface will change
under spread-layer radiation.
The angular rotation velocity $\omega$ of the accreting matter
is known to reach a maximum in the transition zone
between the disk and the spread layer (Shakura and Sunyaev 1973).
Accordingly,
$ d\omega (r=r_n)/dr=0.$
This point in the flow is called the neck (see Figs. 1, 3, and 4).
The dissipative production of thermal energy vanishes at $r_n,$
because the derivative of $\omega (r)$ at this point changes sign
and because the heat release,
which is proportional to $(d\omega/dr)^2,$
is small in the neck zone.
The surface density of viscous energy release in the disk
(in Newton's approximation)
behaves as
$Q^+=$
$(3 \gamma M \dot M/4\pi r^3)(1 - \sqrt{R/r}).$
At $r\gg R,$ the energy release $Q^+$ is a factor of 3 greater
than the local gravitational energy release
$\gamma M \dot M/4\pi r^3,$
which is attributable to the transfer of mechanical energy
by viscous forces over the disk from regions with smaller $r$
to regions with larger $r.$
At the same time, $Q^+ \to 0$ as $r\to R.$
Therefore, the disk half-thickness,
$h=$
$(3/8\pi)(\sigma_T \,\dot M/m_p \, c)(1 - \sqrt{R/r}),$
also decreases as one approaches the neck (Shakura and Sunyaev 1973).
This consideration ignores disk heating by the spread-layer radiation
and the pressure of external radiative flux.
Let us estimate
how the heating by external radiation
affects the disk half-thickness near the neck.
Let $T_s,$ $T_c,$ and $T_r$ be the disk-surface temperature,
the temperature in the disk equatorial plane,
and the external-radiation temperature,
respectively.
Since the flux from the spread layer
accounts for a fraction of the Eddington flux,
its temperature is $T_r = 1-1.5$ keV.
The disk albedo is assumed to be 0.3-0.5.
Consequently, the disk surface near the neck
will be heated to temperatures of the order of 1 keV.
Clearly, the surface temperature 
under quasi-steady-state conditions
(the radial velocity of the matter near the neck is low)
is less than or equal to $T_c.$
Using a hydrostatic estimate of the half-thickness,
$h = R c_s/v_{\varphi}^k,$
and assuming that $c_s \sim \sqrt{T_c/m_p},$
we obtain
$h = 30 R_6^{3/2}\, \sqrt{T_c}/\sqrt{M/M_{\odot}}$ m,
where $T_c$ is in keV,
and $R_6 = R/(10$ km).
It thus follows that the disk angular half-thickness, $\theta_0,$ is finite.
The conclusion about the finite but small disk half-thickness near the neck
is crucial for the validity of our approach
to the problem of accretion onto a neutron star.

Here, we do not consider the advection solution for the flow in the disk
(Narayan and Yi 1995)
for two reasons.
First, the neutron star has a solid surface,
and it is, therefore, difficult to imagine accretion
without energy release on the surface.
Second, the flux from the spread layer is so large
that it will cool the plasma flow
at a distance of several tens of neutron-star radii
through comptonization.
As a result,
an accretion pattern with a relatively thin (geometrically) accretion disk
is most likely realized near the star.

\subsection{Structure of the Spread Layer (Atmosphere)}

We make the following assumptions:

(I) The energy is mainly released in a thin sublayer
near the boundary between the spreading and underlying layers.

(II) Thomson scattering makes a major contribution to the opacity.

(III) We define the free fall acceleration $g_{eff}$
within the spreading layer
with allowance for the contribution of the centrifugal acceleration
as the difference
$g_{eff} = \gamma M/R^2 - v_{\varphi}^2/R - v_{\theta}^2/R$
[see (1.1) and (1.1)'].
For completeness, we added the centrifugal acceleration
due to the meridional motion.
It is small compared to the acceleration $\gamma M/R^2.$
The rotation velocity, $v_{\varphi},$
is assumed to be constant within the atmosphere
(if $\theta$ is fixed).

Under these assumptions,
the atmospheric structure is described by the system of equations
$$
p=g_{eff} \; \Sigma,  \eqno (2.2)
$$$$
q =\frac{c\,\Sigma_T}{3}\;\frac{d\epsilon}{d\Sigma}, \eqno (2.3)
$$$$
p=2\;\frac{\rho\, T}{m_p}+\frac{\epsilon}{3},  \eqno (2.4)
$$$$
\epsilon=a \; T^4.  \eqno (2.5)
$$
The system of equations of hydrostatics (2.2),
radiative heat conduction (2.3),
thermodynamic state (2.4),
and Stefan-Boltzmann's law (2.5)
relates the unknown functions $p, \epsilon, \rho,$ and $T$
of the vertical column density $\Sigma,$
which is a Lagrangian coordinate $(d \Sigma/dy = \rho).$
The $y$ and $\Sigma$ axes are directed downward.
Accordingly, the pressure in (2.2) increases with $\Sigma.$
The $q$ is the absolute flux.
We assume that the layer borders vacuum $(p = 0)$
at $y = 0, \Sigma = 0.$
The boundary-layer bottom
$y = y_S = h,$
$\;\Sigma = \Sigma_S$
is the surface $S$ (see Fig. 4).
The system of equations (2.2)-(2.5)
is written for an arbitrary fixed latitude $\theta.$
Due to the approximate assumption\footnote{
$v_{\varphi}$ is assumed to depend on latitude alone.
By $v_{\varphi}$ we mean the $r$-averaged velocity in the boundary layer.}
of constant $v_{\varphi}(\theta)$ in the boundary layer $0 < y < h,$
the acceleration $g_{eff}$ in (2.2) is constant
as $\Sigma$ increases deep into the layer.

Eq. (2.3) holds for radiative transfer in an optically thick plasma
with $\Sigma_S \gg \Sigma_T$ with dominating Thomson scattering.
We assume that the turbulent dissipation is localized near the surface $S.$
It thus follows that the heat flux $q$ in the spread layer $0 < y < h,$
which is transferred via radiative heat conductance,
is approximately constant.
The total pressure consists of the plasma and radiation pressures.

The system of equations (2.2)-(2.5) can be easily solved.
The solution that satisfies the boundary condition
$$
p(0)=0, \;\;\;
\epsilon(0)=0, \;\;\;
\rho(0)=0, \;\;\;
T(0)=0, \eqno (2.6)
$$
is
$$
p = g_{eff} \Sigma,\;\;\;
\epsilon = \frac{3q}{c}\,\tau_T,\;\;\;
\rho = \frac{m_p\,g_{wr}\,\Sigma_T}{2}\;
\left( \frac{a c}{3q}\;\tau_T^3 \right)^{1/4},\;\;\;
T = \left( \frac{3q}{a c}\;\tau_T \right)^{1/4}, \eqno (2.7)
$$$$
g_{wr}=g_0\, G_{wr},\;\;\;\;
G_{wr} \equiv \Delta =G_{eff}-G_r,\;\;\;\;
G_r=\frac{g_r}{g_0},\;\;\;\;
g_r=\frac{q}{c \,\Sigma_T}, \eqno (2.8)
$$
where $G_{eff}=g_{eff}/g_0$ [see Subsec. 2.2 (III)].
The difference, $\Delta,$ of the gravitational acceleration,
the component of the centrifugal acceleration normal to $S,$
and the radiation acceleration $g_r$ (2.8)
is particularly important for the subsequent discussion.

Let us derive the dependence on $y.$
The Lagrangian and Eulerian differentials
are related by $d\Sigma = \rho \, dy.$
Substituting the solutions (2.7) and (2.8) into this relation
and integrating it using the conditions (2.6),
we obtain
$$
\Sigma = \frac{\beta \; \Delta^4}{3\times 2^{12}\, q}\, y^4 =
\frac{y\,\rho}{4}, \;\;\;
\rho = \frac{\beta\;\Delta^4}{3\times 2^{10}\, q}\,y^3,\;\;\;
\beta = m_p^4\;g_0^4\;a\,c\,\Sigma_T, \eqno (2.9)
$$$$
p_{pl} = g_{wr}\;\Sigma,\;\;\;
p_r=g_r\,\Sigma,\;\;\;
p=p_{pl}+p_r = g_{eff}\,\Sigma,\;\;\;
T=\frac{m_p\;g_{wr}}{8}\,y. \eqno (2.10)
$$
These solutions are very simple.
The atmosphere under consideration is similar in structure
to the atmospheres of X-ray bursters during outbursts
and to the atmospheres of supermassive stars.
We describe these solutions in some length,
because the calculations of transport characteristics
are based on them.

\subsection{Lateral Force and the Surface Energy Density}

Here we study the dynamics of a layer of shallow water or a fluid film
that flows over the underlying surface $S$ (see Fig. 4).
The thickness, $h,$ of the layer is assumed to be small
compared to its horizontal extent $R \theta$ [see (2.1)].
When the equations with gradients in $\theta$ are derived,
the layer is broken up into the differential elements
$R \,\delta \theta.$
The force interaction between adjacent elements
is effected by the lateral force
$\int_0^h\, p\,dy.$
Substituting (2.9) and (2.10), we obtain
$$
\int_0^h\;p \,dy =
\frac{8}{5}\;\frac{G_{eff}}{\Delta}\;\frac{T_S\; \Sigma_S}{m_p},
\eqno (2.11)
$$
where, as above,
the subscript $S$ denotes values near the bottom of the spread layer.
It is important that the force (2.11)
contains the large dimensionless factor $G_{eff}/\Delta$
at $\Delta \ll 1.$

The layer transports mass, momentum, angular momentum, and energy
over the spherical surface $S.$
In particular, the advection of radiative energy takes place.
The expressions for the surface densities of mass,
momentum, and angular momentum are obvious:
$\Sigma_S,$
$\;v_{\theta}\, \Sigma_S,$
$R\,\cos\theta \, v_{\varphi}\,\Sigma_S,$
where $v_{\theta}(\theta)$ is the $r$-averaged meridional velocity
in the layer,
and $R \cos \theta$ is the arm relative to the rotation axis,
which the polar axis is.
Let us calculate the surface energy density (erg cm$^{ - 2 })$
using the distributions (2.9) and (2.10).
The total energy
of the layer consists of the kinetic energy
$
K = (v_{\theta}^2 + v_{\varphi}^2) \, \Sigma_S/2,
$
the gravitational energy $(E_g),$
and the internal energy $(E_{int}).$
In turn, the latter consists of the plasma $(E_{pl})$ and radiative $(E_r)$
contributions.
The surface density of gravitational energy is
$$
E_g = \int_0^h\;\rho\,g_{eff}\, (h-y) dy=
\frac{8}{5}\;\frac{G_{eff}}{\Delta}\;\frac{T_S\;\Sigma_S}{m_p}. \eqno (2.12)
$$
Interestingly, the expressions (2.11) and (2.12) are identical.
The energy (2.12) is measured
 from the bottom of the layer $y = h.$
Since the volume density of plasma internal energy is
$3 \rho T/m_p = (3/2)p_{pl},$
we have 
$E_{pl}=(3/2) \int_0^h \; p_{pl}\, dy.$
Calculating this integral and the integral
$$
E_r = \int_0^h\; \epsilon\,dy =
(24/5)\,(G_r/\Delta)\,T_S \, \Sigma_S/m_p,
$$
we obtain
$$
E_{int}=\frac{12}{5}\;\frac{G_{eff}+G_r}{\Delta}\;
 \frac{T_S\;\Sigma_S}{m_p}. \eqno (2.13)
$$
All expressions (2.11)-(2.13)
that appear in the formulas for horizontal interactions
in the layer
increase as the layer swells with $\Delta\to 0.$

\subsection{Radiative Cooling}

The energy losses by the layer surface are determined by the flux
which comes from the inside
and can be readily expressed in terms of the layer surface temperature, $T_v,$
on the side of vacuum
$(q = a\, c\, T_v^4/4).$
This is the temperature at an optical depth of the order of unity
$(\Sigma_v \simeq \Sigma_T).$
The solutions (2.7) and (2.8),
in which the integration constant
for the equation of radiative heat transfer (2.3)
was discarded,
is valid at large $\Sigma$ $\;(\Sigma \gg \Sigma_T).$
Formula (2.7) for $T$ allows the local flux, $q,$ to be expressed in terms
of the local $\Sigma$ and $T$
$\;(q = a\,c\,T^4/3\,\tau_T).$
In particular, inside the spread layer near its bottom,
we obtain
$$
q = a \, c \, T_S^4 \; \Sigma_T/3 \, \Sigma_S. \eqno (2.14)
$$

\subsection{Turbulent Deceleration of Rotation and Viscous Energy Release}

{\bf Profiles of average and fluctuation velocities.}
Let us dwell on friction and dissipation
and analyze the friction of the boundary layer against the star.
The turbulence via the friction
it produced
decelerates and heats up the layer.
In this case, the rotation and meridional spread of the matter in the layer
draws the underlying layers into motion
through turbulent viscous gripping.
This issue is discussed in Subsec. 2.1.
Since the source of heat concentrates near the boundary $S,$
the temperature under the surface $S$ is constant,
while the entropy $s$ rapidly decreases with depth.
Stratification with a large reserve of gravitational stability arises
(the Richardson number is great in the underlying layers).
It is difficult for the high-entropy layer,
which moves from above,
 to "grip" the low-entropy underlying layer.
This situation roughly corresponds to a wind above a smooth surface.

{\it Turbulence in a gaseous layer}
that flows with a subsonic velocity above a fixed boundary
has been well studied and described by 
Loitsyanskii (1973),
Schlichting (1965),
Chugaev (1982),
Landau and Lifshitz (1986),
and, recently, by Lin et al. (1997).
For low viscosity,
the height distribution of the mean velocity
is described by Prandtl-Karman's universal logarithmic profile
$$
\langle v \rangle \approx \frac{5}{2} \, v_{\star}\,
\ln (7.7 \;v_{\star}\; \Delta y/\nu),\;\;\;\;\;\;\;
\tau = \rho v_{\star}^2,  \eqno (2.15)
$$
where $\Delta y$ is {\it measured from the surface} $S,$
and $\tau$ is the tangential stress
which does not depend on $\Delta y.$
The parameter $v_{\star}$ is important.
It determines the turbulent velocity and pressure pulsations.
The universal profile is such
that $v_{\star}$ is velocity pulsations about the mean in all directions:
$v_{\star}\simeq$
$\sqrt{\, \langle (v_{\theta}- \langle v_{\theta} \rangle )^2 \rangle }\simeq$
$\sqrt{\langle v_r^2 \rangle }\simeq$
$\sqrt{\, \langle (v_{\varphi}- \langle v_{\varphi} \rangle )^2 \rangle}$
or
$$
v_{\star}=
\sqrt{
 \,\langle\,(\vec v-
\langle v_{\varphi}\rangle \, \vec e_z -
\langle v_{\theta}\rangle \, \vec e_x \,)^2 \,\rangle\,/3 },
$$
where $\vec v = \{v_{\theta}, v_r, v_{\varphi}\},$
and $x, z$ is the wall plane.
This implies that the pulsations are isotropic
at a given distance, $\Delta y,$ from the wall
in a comoving frame of reference
which moves with the mean velocity $\langle v \rangle$ of the flow
at this distance.
The order-of-magnitude amplitude of the pressure pulsations about the mean
is $\rho v_{\star}^2.$
An important point is that
the velocity $v_{\star}$ and the tangential stress $\tau$
in the region of turbulent mixing
do not vary as the distance to the wall $\Delta y$ is varied.
Actually, the law (2.15) follows from the condition
of the invariance of $\tau.$
When this law is deduced,
the stress is written in Newton's form
$\tau = \rho \, \nu_t \, d \langle v \rangle /dy,$
and the mixing-length theory is used,
in which the turbulent viscosity $\nu_t = l_t\, v_{\star}$
is determined by the scale of the wall vortices
$l_t = \kappa \, \Delta y.$
The coefficient $\kappa\approx 0.4$ is called Karman's coefficient.
The integration constant in Newton's friction law
$\langle v\rangle = \int dy \,\tau \dots,$
which appears in (2.5) under the logarithm,
is chosen by joining the viscous and turbulent solutions
in the transition region between the viscous sublayer and the turbulent region.
Formula (2.15) holds outside the viscous sublayer,
when $\Delta y > l_{\nu},$
$\; v_{\star} \, l_{\nu}/\nu \approx 12,$
where $l_{\nu}$ is the thickness of the viscous sublayer
which is determined by the molecular and radiative viscosities.

The disk-accretion theory commonly uses an expression
for the kinematic viscosity $\nu_t$
of the form $\nu_t = \alpha \, h_d \,(c_s)_d,$
where $h_d$ and $(c_s)_d$ are the disk thickness and the speed of sound
in the disk, respectively.
For the Keplerian dependence of angular velocity
$\omega_d$ on $r,$
this expression reduces to a simple formula for the stress
$\tau = \alpha p,$
where $p$ is the pressure in the disk (Shakura and Sunyaev 1973).
Let us compare the formulas
$    \nu_t= \alpha \, h_d \, (c_s)_d$
and $\nu_t= \kappa \, \Delta y \, v_{\star}.$
We see that the wall turbulence differs from the disk turbulence,
first, by the form in which the mixing length is written
$(\kappa \, \Delta y$ instead of $h_d)$
and, second, by the formula for turbulent velocity pulsations
$[v_{\star}$ instead of $\alpha \, (c_s)_d].$

{\bf Friction coefficient.}
Let us calculate the stress $\tau.$
For this purpose, we extend the distribution (2.15),
which is weakly dependent on $\Delta y,$
to the height $\Delta y = h.$
To obtain an estimate,
we assume that at this height
the mean velocity of flow $\langle v \rangle$
is equal to the Keplerian velocity.
The ion viscosity is
$\nu_i=$
$2.2 \times 10^{-15} \; T^{2.5}/\rho_S \; \ln \Lambda$
$=1.3 \times 10^4$ cm$^2$ s$^{-1},$
where,
$T$ is in degrees,
and $\rho_S$ is in g cm$^{-3}$ (Spitzer 1962).
We also assume the following:
$h = 100$ m,
$\Sigma_S = 300$ g cm$^{-2},$ $T = 3$ keV,
and the Coulomb logarithm $\ln \Lambda = 10.$
This viscosity is small compared to the {\it radiative viscosity}
$$
\nu_r = \frac{4}{15} \; \frac{a \,T^4\, \Sigma_T}{c\,\rho^2}
\simeq \frac{\rho_r}{\rho} \, l_T \; c
$$
(Weinberg 1972);
here, $\rho_r = a T^4/c^2$ is the photon "density",
$l_T = 1/n \sigma_T$ is the photon mean free path,
and $n = \rho /m_p.$
It follows from the numerical calculations
that the radiation density $\rho_r$
accounts for a small fraction $(\sim 10^{-2})$
of the plasma density $\rho_S.$
For the above values of $\rho$ and $T,$
we obtain $\nu_r = 1.7 \times 10^7$ cm$^2$ s$^{-1}.$
The mixing is governed by the radiative viscosity:
$\nu_r \gg \nu_i.$
Substituting $\nu_r$ in (2.15)
and setting $M = 1.4 M_{\odot},$
and $R = 12$ km
when calculating the Keplerian velocity,
we arrive at the equation
$4.2 \times 10^9 = v_{\star} \ln (v_{\star}/220),$
where the unknown is in cm s$^{-1}.$
The estimates are given in the approximation of Newton's potential.
Solving this equation,
we obtain
$v_{\star} \approx 3\times 10^8$ cm s$^{-1}.$
In this case, the viscous-layer thickness
$l_{\nu} \approx$
$12 \nu/v_{\star} \approx$
$0.7$ cm is much smaller than the scale height $h.$
The Reynolds number is
Re$ = h v_{\varphi}^k/\nu \sim 10^{7}.$
Let us write
$$
\tau = \alpha_b \; \rho \; v^2. \eqno (2.16)
$$
The coefficient of friction against the bottom
is $\alpha_b = v_{\star}^2/v^2$
[cf. (2.15) and (2.16)].
In our example,
$\alpha_b \approx (v_{\star}/v_{\varphi}^k)^2 \approx 0.81 \times 10^{-3}.$
The value of $\alpha_b$ changes only slightly
as the parameters are varied.
In Subsec. 3.1, we compare the terms
which are associated with $\alpha_b$ (wall turbulence)
and $\alpha$
(a coefficient that is widely used in the disk-accretion theory).

{\bf Dissipation and the height distribution of energy release.}
The volume power density (erg s$^{-1}$ cm$^{-3})$
of the heat source $\dot \epsilon_t$ attributable to turbulent friction
is given by
$$
\dot \epsilon_t (\Delta y) \approx
(5/2) \; v_{\star} \; \tau / \Delta y \eqno (2.17)
$$
(Landau and Lifshitz 1986).
It follows from (2.17) that the surface density
of the total viscous energy release is
$$
Q^+ = \int_0^h \, \dot \epsilon_t (\Delta y) \;d (\Delta y)\approx
\int_{l_{\nu}}^h \, \dot \epsilon_t \;d (\Delta y) \approx
2.5 \,\tau \,v_{\star}\; \ln \frac{h}{l_{\nu}}\approx \tau v. \eqno (2.18)
$$
Here, $v$ is the height-averaged velocity in the layer.
In the case under consideration,
where the flow moves latitudinally at velocity $v_{\varphi}$
and meridionally at velocity $v_{\theta},$
the total velocity
$\sqrt{ v_{\theta}^2 + v_{\varphi}^2 }$
must be substituted in (2.18).
The energy release (2.17)
takes place mainly near the bottom of the boundary layer.
For example,
$$
[\, \int_{0.1 h}^h \, \dot \epsilon_t \;d (\Delta y)\,] / Q^+
\approx 0.2.
$$
An analysis of the calculations
which are given below
shows that $v_{\varphi}$ is several times faster
than the speed of sound $c_s$ at $\theta < \theta_{\star}$ --
{\it the flow is moderately supersonic.}
In this case, the expressions (2.15)-(2.18) {\it remain valid}
(Loitsyanskii 1973; Schlichting 1965).
In a supersonic case with the Mach number
$\Ma \simeq 5,$
$\alpha_b$ in (2.16) decreases by a factor of 1.5 to 2.

Above, we described the surface characteristics (Subsec. 2.3),
radiative cooling (Subsec. 2.4),
and dissipative heating (Subsec. 2.5).
Now, we can proceed to a derivation of the balance equations
that relates to all these physical processes.

\subsection{Equations of mass, momentum, and angular momentum transfer}

Let us derive the first three equations
of the system of four $1D$ equations
which represent the laws of conservation of mass
(equation $\Sigma),$
momentum (or angular momentum along the meridian) in $\theta$
(equation $v_{\theta}),$
angular momentum (equation $v_{\varphi}),$
and energy (equation $T).$
We take a film annulus
with the area $dS = 2 \pi R C \, R \delta\theta$
on the sphere $S,$
where $C = \cos\theta.$
Its mass is $dS \;\Sigma_S.$
The mass flux through the $\theta = \const$ conic surface
(annulus boundary)
is
$2 \pi R C v_{\theta} \Sigma_S.$
Accordingly, the equation $(\Sigma)$ is
$$
R C (\Sigma_S)'_t = - (C \Sigma_S \; v_{\theta})'_{\theta}.  \eqno (2.19)
$$

{\bf Equation $(v_{\theta}).$}
The momentum flux is
$2 \pi R C v_{\theta} \Sigma_S \; v_{\theta},$
the lateral force is
$2 \pi R C \int_0^h p dy,$
and the component of the centrifugal force
which is tangential to the sphere is
$(dS\;\Sigma_S\; v_{\varphi}^2/RC)\sin \theta.$
The meridional friction force $\tau_{\theta} \; dS$
is
given by the relations
$$
\tau=\alpha_b \;\rho_S \; v^2,\;
v=\sqrt{v_{\theta}^2+v_{\varphi}^2},\;\,
\tau_{\theta} = \tau \sin \beta,\;
\sin\beta=v_{\theta}/v,\;
\tau_{\theta}=\alpha_b \;\rho_S \; v_{\theta} \,v,
$$
where $\tau_{\theta}$ is the meridional component
of the total shear stress,
the vector $\vec \tau$ is directed opposite
to the total flow velocity vector
$v_{\theta} \vec e_{\theta} + v_{\varphi} \vec e_{\varphi},$
$\rho_S$ is the density at the bottom of the spread layer,
and $\beta$ is the angle
between the total velocity vector and the latitude.
These relations follow from (2.16).
Substituting (2.11) into the momentum equation, we obtain
$$
R C (\Sigma_S\; v_{\theta})'_t = - (C \Sigma_S\; v_{\theta}^2)_{\theta} -
\frac{8}{5} C \left(\frac{G_{eff}}{\Delta}\frac{T_S\Sigma_S}{m_p}\right)'_{\theta}
- R C \tau_{\theta} -
\sin\theta \Sigma_S \; v_{\varphi}^2.  \eqno (2.20)
$$
The bottom density $\rho_S$ is related to the sought-for functions
$v_{\theta}, v_{\varphi}$ and $T_S$
(below, we eliminate $\Sigma_S$ by using the mass integral)
by
$$
\rho_S=\frac{4\,\Sigma_S}{h}=\frac{m_p \, g_{wr} \,\Sigma_S}{2 \,T_S},
\;\;\;\;\;\;\;\;\;
h=\frac{8 T_S}{m_p\, g_0 \,\Delta},
\eqno (2.21)
$$
where $q_{wr}$ is given by (2.8).
We see that at the same surface density of the matter,
the {\it layer thickness,} $h,$ is inversely proportional to $\Delta.$
The smaller $\Delta,$ the higher the accuracy
with which the radiation-pressure force offsets the difference
between the gravitational and centrifugal forces.

The subsystem (2.19), (2.20) resembles the $1D$ hydrodynamic equations
$\rho'_t = (\rho u)'_x,$
$\;\rho u'_t = - \rho u u'_x - (\rho T/m_p)'_x.$
Note the "pressure" enhancement in (2.20)
compared to the formula for an "ideal gas" $T_S\;\Sigma_S$
because of the factor $G_{eff}/\Delta$
(see similar remarks in Subsec. 2.3).

{\bf Equation $(v_{\varphi}).$}
The rate of change in the angular momentum of the boundary-layer annulus
is
$dS\,(\Sigma_S\;v_{\varphi})'_t \,R C.$
Writing the angular momentum flux
$ 2 \pi R \, C v_{\theta} \, \Sigma_S \; v_{\varphi} R \, C$
and the friction torque about the polar axis as
$$
\tau_{\varphi} \, dS \, R\, C, \;\;\;\;\;\;\;
\tau_{\varphi} = \alpha_b \;\rho_S \; v_{\varphi} \, \sqrt{v_{\theta}^2+v_{\varphi}^2},
$$
we obtain
$$
R\, C^2 \,(\Sigma_S\; v_{\varphi})'_t = - (C^2 \;\Sigma_S\; v_{\theta} \,v_{\varphi})'_{\theta} -
R \,C^2 \;\alpha_b \;\rho_S \;v_{\varphi} \,\sqrt{v_{\theta}^2+v_{\varphi}^2}. \eqno (2.22)
$$
In (2.22), in order to determine the latitudinal friction $\tau_{\varphi},$
we projected the total stress $\tau$ onto the latitude,
much as was done above
when the meridional friction was calculated.

\subsection{Energy Transfer and Emission by the Layer}

Let us write out the {\it energy equation} $(T).$
In this equation the formulas
for the rate of change in the total energy of the annulus,
for the enthalpy flux
(the total energy flux and the power of lateral forces),
and for the radiative losses
are used.
They are
$$
dS \;\left(K+E_{sum}\right)'_t,\;\;\;\;\;
E_{sum} = E_g + E_{int} =
\frac{4 T_S\;\Sigma_S}{5 m_p}\;\;\frac{5 G_{eff}+3 G_r}{\Delta},
$$
$$
2 \pi R C v_{\theta} \, (K+H),\;\;\;\;\;
H = E_{sum} + \int_0^h \, p dy = \frac{4 T_S\;\Sigma_S}{5 m_p}\;
\frac{7 G_{eff}+3 G_r}{\Delta}
$$
$$
q\, dS,\;\;\;\;\;\;\;K=(1/2) \, \Sigma_S\;(v_{\theta}^2+v_{\varphi}^2).
$$
Adding up these expressions,
we obtain the sought-for equation for the total energy
$$
R C \left(E_{sum}+ K\right)'_t =
-\left[C v_{\theta} \left(H+K\right)\right]'_{\theta}-R C q. \eqno (2.23)
$$
This equation allows for the energy (including radiative energy) advection
and radiative cooling.

\section{TRANSFORMATIONS AND ANALYSIS OF THE SYSTEM}

\subsection{Steady-State Spread}

In the steady-state case, Eq. (2.19) is integrable.
From the condition of {\it mass conservation,} we have
$$
\frac{1}{2}\;\dot M= 2\pi R C v_{\theta} \Sigma_S \eqno (3.1)
$$
i.e., the matter spreads out over the two hemispheres.
Since $C \,v_{\theta} \,\Sigma_S,$ is constant,
from the system (2.20), (2.22), and (2.23) we derive
$$
C\, \Sigma_S \; v_{\theta}\; v_{\theta}'+\frac{4}{5}\, C \left(\frac{G_{eff}}{\Delta}\,
 \frac{2 \,T_S}{m_p}\,
\Sigma_S \right)' =
 - R \,C \,\tau_{\theta} - \sin \theta \,\Sigma_S \; v_{\varphi}^2,
\eqno (3.2)
$$$$
\tau_{\theta}=\alpha_b \;\rho_S \;v_{\theta} \,\sqrt{v_{\theta}^2+v_{\varphi}^2},
$$$$
C \,\Sigma_S \; v_{\theta} \,(C \,v_{\varphi})' = - R\, C^2 \,\tau_{\varphi},\;\;\;\;\;\;\;
\tau_{\varphi}=\alpha_b \;\rho_S\; v_{\varphi} \,\sqrt{v_{\theta}^2+v_{\varphi}^2}, \eqno (3.3)
$$$$
C\, \Sigma_S \; v_{\theta} \left( \frac{v_{\theta}^2+v_{\varphi}^2}{2}+
\frac{2}{5}\;\frac{2 T_S}{m_p}\;\frac{7G_{eff}+3G_r}{\Delta} \right)'
= - R \,C \,q.  \eqno (3.4)
$$
The prime denotes differentiation with respect to $\theta.$
Because of the integral (3.1),
the number of unknowns decreases.
Eqs. (2.23) and (3.4) reflect the absence of radiative flux
from the layer to the star.
In addition,
we ignore the mechanical work between the layer and the star
$(\tau v_{\varphi}^{star}\approx 0),$
because the star is massive 
$(\rho^{star}$ is great even in the underlying layer)
and because the velocity of its surface layers is low
$(v_{\varphi}^{star}\ll v_{\varphi};$
see Subsecs. 2.1 and 2.5).

{\bf Comparison with $\alpha-$friction.}
{\it We disregard the friction between adjacent annuli}
(or differential elements $\delta \theta)$
in comparison with the friction against the bottom.
Indeed, in the theory of disk accretion (Shakura and Sunyaev 1973)
with $"\alpha$-friction",
the turbulent stress $\tau$
between the annuli is
$\alpha \, h \, c_s \, \rho \; (dv_{\varphi} / dx),$
where $h$ is the layer thickness (see Subsec. 2.1),
$h c_s$ gives an estimate
of the maximum possible hydrodynamic interaction,
$\alpha$ is a coefficient that is small compared to unity,
and $x$ is the arc length along the meridian.
If we take into account this interaction,
instead of Eq. (3.3) we obtain,
for example,
$$
- C \,v_{\theta} \,\Sigma_S\;(C\, v_{\varphi})' -
R\, C^2 \;\tau_{\varphi} +
(\alpha/ R) \,(C^2 \,h^2\, \rho\, c_s \, v_{\varphi}')' = 0.
$$
This equation contains the second-order derivative
with respect to $v_{\varphi}.$
The ratio of the third and second terms is
$$
(\alpha/\alpha_b)\,(c_s/v_{\varphi})\,(h/R\theta_{\star})^2.
$$
Neglect of the friction between adjacent annuli
is justified by the smallness of
$(h/R\theta_{\star})^2$
in the shallow-water approximation (2.1).
Note that the term with
$\tau_{\varphi}$
drops out in the angular momentum equation for an accretion disk --
there is no friction against the bottom.

If we set $C\approx 1$ in Eq. (3.4),
discard $v_{\theta}^2$
compared to
$v_{\varphi}^2,$
differentiate $v_{\varphi}^2,$
use (3.3) to eliminate $v_{\varphi}',$
and ignore radiative losses $q,$
then we obtain
$(v_{\theta} \,H)' = R \, \tau_{\varphi} \, v_{\varphi}.$
This equation together with Eq. (3.3)
describe the transformation of rotational energy into enthalpy.

\subsection{Natural Scales}

Let us rewrite the system (3.2)-(3.4) in dimensionless form.
In units of the stellar radius $R = 1$
and the Keplerian velocity at the stellar equator $v_{\varphi}^k=1,$
it takes the form
$$
U^2 U'+ \frac{4}{5} \,C\,U^2 \left(\frac{G_{eff}}{\Delta}\;\frac{\hat T}{C\,U}\right)'=
- F_{\theta} - F_{cf}, \;\;\;\;\; C\equiv \cos \theta\eqno (3.5)
$$$$
U W W' = - F_{\varphi} + F_{cf}, \eqno (3.6)
$$$$
U^2 U'+\frac{2}{5} U \left(\frac{7G_{eff}+3G_r}{\Delta} \hat T \right)' =
F_{\varphi} - F_{cf} - \dot E_{rad}. \eqno (3.7)
$$
In what follows,
$U=v_{\theta}/v_{\varphi}^k,$
$W=v_{\varphi}/v_{\varphi}^k,$
 and
$\hat T=2T_S/m_p \,(v_{\varphi}^k)^2$
 are the dimensionless sought-for functions.
 Eq. (3.5) can be derived from (3.2) by multiplying it by
$v_{\theta}.$
 Below, we give the functions and the notation
 that are used to write the system.
 The function
$$
\Delta=\Delta (U,W,\hat T;\theta)= 1-U^2-W^2-G_r,\;\;\;\;\;\;\;
G_r = B C U \hat T^4, \eqno (3.8)
$$$$
B=\frac{4\pi}{3}R^2\;\frac{a (m_p (v_{\varphi}^k)^2/2)^4}{\dot M v_{\varphi}^k}=
\frac{0.65\times 10^{18}(M/M_{\odot})^{7/2} }{
R_6^{5/2}\,(L_{SL}/L_{edd})} \eqno (3.9)
$$
[see (1.2), (2.8), (2.14), and (3.1)].
 For completeness,
 we included the centrifugal acceleration
 which is associated with the meridional motion in expression (3.8) for
$\Delta.$
 In the radiating layer,
 it is small compared to the acceleration
 which is associated with the latitudinal motion:
$ U^2 \ll W^2.$
 In the dark layer, the centrifugal forces are insignificant.
 In (3.9) and below
$\dot M = 10^{17} \, \dot M_{17}$ g s$^{-1}.$
The terms
$F_{\theta},$ $F_{\varphi},$ $F_{cf},$ and $\dot E_{rad}$
 which are related to the friction, 
the centrifugal force (cf),
 and the radiation are given by
$$
F_{\theta}=\frac{\alpha_b U^2 V \Delta}{\hat T},\,\;\;\;
F_{\varphi}=\frac{\alpha_b W^2 V \Delta}{\hat T},\,\;\;\;
F_{cf}=\frac{S}{C} U W^2, \,\;\;\;
\dot E_{rad} = A C^2 U^2 \hat T^4. \eqno (3.10)
$$$$
A=\frac{1}{2}\; \frac{B}{(L_{SL}/L_{edd})},\;\;\;\;\;
S\equiv\sin\theta,\;\;\;\;C\equiv\cos\theta,\;\;\;\;
V=\sqrt{U^2+W^2}.
\eqno (3.11)
$$
We see that the decrease in $\Delta$
leads to a decrease in $F_{\theta}$ and $F_{\varphi}$
[see also (2.20)-(2.22)].
 Physically, this is caused by the drop in density (2.21)
 at the bottom due to an increase in the boundary-layer thickness.

The system of equations (3.5)-(3.7)
 for the sought-for functions $U, W,$ and $\hat T$
 includes three parameters, $L_{SL}/L_{edd},$ $B,$ and $\alpha_b.$

\subsection{Transformation to an Explicit Form}

The system (3.5)-(3.7) relates the derivatives of the unknown functions
$U, W,$ and $\hat T.$
 For this system to be numerically integrated,
it must be resolved for the derivatives.
 The functions $U, W,$ and $\hat T$
 enter into the pressure and the enthalpy
 which are differentiated in Eqs. (3.5) and (3.7)
 in a complicated way via the functions
$\Delta, G_{eff,}$ and $G_r.$
Since the transformation of these equations to a form
 that contains only the derivatives of the sought-for functions
 turns out to be cumbersome, 
it makes sense to give the final results of this transformation.
 We have
$$
U'   =(   e_{\tau} u_{00}-u_{\tau} e_{00})/d, \eqno (3.12)
$$ $$
W'=\omega,\;\;\;\;\;\;\; \omega= ( - F_{\varphi} + F_{cf} )/U W, \eqno (3.13)
$$ $$
\hat T'=( - e_{u   } u_{00}+u_{u   } e_{00})/d, \eqno (3.14)
$$
where
$$
u_u=U^2-\frac{4}{5}\;
\frac{ G_{eff}^2 - 2 G_{eff} G_r - 2 U^2 G_r }{\Delta^2}
\;\hat T,\;\;\;\;\;\;\;
u_{\tau}=\frac{4}{5}\frac{G_{eff} (G_{eff}+3G_r)}{\Delta^2}U,
$$$$
u_w=\frac{8 G_r U W \hat T}{5\Delta^2},\;\;\;\;\;\;\;
u_0= - \frac{4}{5}\frac{\sin\theta}{C}
\frac{G_{eff} ( - G_{eff} + 2 G_r) }{\Delta^2}U\hat T,
$$$$
e_u=U^2 + 4 \frac{ G_r (G_{eff} + 2 U^2) }{\Delta^2}\hat T,\;\;\;\;\;\;\;
e_{\tau}=\frac{2}{5}\;\,\frac{ 7 G_{eff}^2 + 36 G_{eff} G_r - 3 G_r^2 }{\Delta^2} U,
$$$$
e_w= 8 \frac{G_r}{\Delta^2} U W \hat T,\;\;\;\;\;\;\;
e_0= - 4\frac{\sin\theta}{C}\frac{G_{eff} G_r}{\Delta^2} U\hat T,
$$$$
d=u_u e_{\tau}-u_{\tau}e_u, \eqno (3.15)
$$$$
u_{00}=-u_0-u_w\omega-F_{\theta}-F_{cf},\;\;\;\;\;\;
e_{00}=-e_0-e_w\omega+F_w-F_{cf}-\dot E_{rad}.
$$
Here, $u_u, u_w, u_{\tau}, e_u, e_w,$ and $e_{\tau}$
 are the coefficients at the derivatives
$U', W',$ and $\hat T'$ in Eqs. (3.5) $(u)$
 and (3.7) $(e),$
respectively.
 We represented Eqs. (3.5) and (3.7) as
$u_u U' + u_{\tau} \hat T' = u_{00}$
 and 
$e_u U' + e_{\tau} \hat T' = e_{00}$
 and then solved this linear system for $U'$ and $T'.$

\subsection{Phase Space and its Section by the 'Levitation' Surface
 $(\Delta = 0)$}

The four-dimensional
$(\theta, U, W, \hat T)$
 phase space of the system 
(3.12)-(3.14) is filled with the integral curves
$U(\theta), W(\theta), \hat T(\theta).$
 Each "trajectory" (along $\theta)$ of this stream of integral curves
 represents a steady flow of the accreting plasma
 over the stellar surface $S$ (Figs. 2-4)
 which locally satisfies
 the differential balances of mass, momentum, angular momentum, and energy
 (2.19), (2.20), (2.22), and (2.23).
 It is necessary to study and classify all these flows.

An analysis shows
 that the numerical integration is stable only in the direction of increasing $\theta.$
 The initial data
 $\theta_0$ and $U_0=U(\theta_0), W_0=W(\theta_0), \hat T_0=\hat T(\theta_0)$
 at the interface between the disk and the spread layer are therefore required.
 There are many initial points in the four-dimensional set
$\theta_0, U_0, W_0,$ and $\hat T_0.$ 
Here, $\theta_0$ is the angular width of the disk
 in the zone of its contact with the layer;
$ U_0$ and $W_0$ are the initial spread and rotation velocities, respectively;
 and $\hat T_0$ is the initial temperature.
 The first narrowing of the exhaustive search
 for
$\theta_0, U_0, W_0,$ and $\hat T_0$ 
 is ensured by the fact that $\theta_0$
turns out to be insignificant if
$\theta_0\ll\theta_{\star}.$
 The second narrowing is associated with the separation
 of
 $W_0$ from the set of initial data.
 At the disk-layer interface, $W$ is close to the Keplerian velocity.
 Accordingly, we set $W_0 = 1 - \delta$
 and perform three series of calculations
 with $\delta = 10^{ - 1}, 10^{ - 2},$ and $10^{ - 3.}$
 In each series, the initial points lie on the $U_0, T_0$
 (spread velocity, temperature at the layer bottom)
 plane.

A special place in the classification analysis is occupied by the 
$\Delta (\theta, U, W, \hat T) = 0$ 
"levitation" surface of the accretion layer,
 where $\Delta$ is a combination
 of the gravitational and centrifugal forces and the radiation-pressure force
 [see (2.8) and (3.8)].
 Let us consider the intersection
of this surface
 with $(U, \hat T)$ planes,
 in particular,
 with the $(U_0, \hat T_0)$ plane.
 In each section, the $\Delta = 0$ curve is a hyperbola,
$$
U \hat T^4=\const  \eqno (3.16)
$$
(we assume that $\theta$ is small
 and that $C = \cos \theta \approx 1);$ see (2.8) and (3.8).
 There are the regions $\Delta<0$ and $\Delta>0$ 
above and under the hyperbola (3.16).
 Clearly, only the region of initial data under the hyperbola is physically meaningful,
 because in this case,
 the total acceleration $g_{wr}$ is directed downward
 and because the gravitation presses the layer against the stellar surface.

\subsection{Attracting or Limiting Surface and One-Parameter Structure}

Let us consider the "triangular" area $\Delta > 0$
 that is bound by the $U_0$ and $\hat T_0$ axes
 and by the hyperbola (3.16).
This area contains the boundary $(\Delta \ll 1)$
 and distant $(\Delta \sim G_{eff})$ subregions.
These subregions differ greatly in area.
 Accordingly, the trajectory of the common position
 starts from the subregion $\Delta \sim G_{eff}.$
 Consider Eqs. (3.5)-(3.7).
 An important point is that $U_0$ and $\hat T_0$ are small.
 This is the reason why 
$|d\ln\hat T/d\theta|$$\gg $$1$
 in the distant subregion
 and 
 $|d\ln\hat T/d\theta|$$\sim $$1$
 in the boundary subregion,
 implying that the "hard" or steep segments of the trajectories
 lie in the subregion
 which is farther from the $\Delta = 0$ surface (deep in the "triangular" area) 
and the "soft" or gentle segments lie in the boundary subregion.

Let us study the general trajectory
 on which $\Delta \sim G_{eff}$ at the starting point.
 In the immediate vicinity of the starting point
 (i.e., on a steep segment),
 we can assume for a rough estimate that the basic equation
 is the energy equation (3.7).
$\Delta$ changes most rapidly during the displacement in $\theta,$
 because it contains the fourth power of $\hat T.$
 The remaining variables $(U, W, \hat T)$ change more slowly.
 We disregard their change and factor them outside the differentiation. 
Thus, only one unknown function $\hat T^4$ remains.
 Assume that $C \approx 1$ and $G_r \simeq G_{eff}.$
 In the right part of the energy equation,
 we discard the small power $F_{cf}$ of the centrifugal force
 compared to the power $F_{\varphi}$
 which is associated with the latitudinal friction stress
$\tau_{\varphi}$
 and radiative losses.
 Taking 
$\Delta\approx 1-W^2-B U \hat T^4$ 
as the unknown function 
instead of $\hat T^4$ and rewriting the equation for this new unknown function, 
we obtain
$$
\frac{\Delta'}{\Delta^2}= - \frac{\alpha W^3/\hat T+(A/B) U}{4 G_{eff} \hat T U}\Delta+
\frac{(A/B)}{4\hat T}. \eqno (3.17)
$$
We substitute the following typical values:
$ A/B \sim 10,$
$\, U \sim 10^{ - 5},$
$\, \hat T \sim 10^{-4},$
$\, W=1-\delta,$
$\, 10^{-3}<\delta <10^{-1},$
$\, \alpha_b\sim 10^{-3}.$
It follows from the estimate
 that the first term (friction) dominates in the numerator of the fraction
 before $\Delta$ in (3.17).
For $\Delta \sim G_{eff},$
 the first term dominates in the right part of (3.17).
 We see that $\Delta$ rapidly decreases on the steep segment.
This decrease causes $\Delta$ to drop to such a small value
 that the first and second terms in the right part of (3.17) become equal.

\begin{figure}[tbh]
\epsfxsize=8cm
\epsfbox[-200 185 400 656]{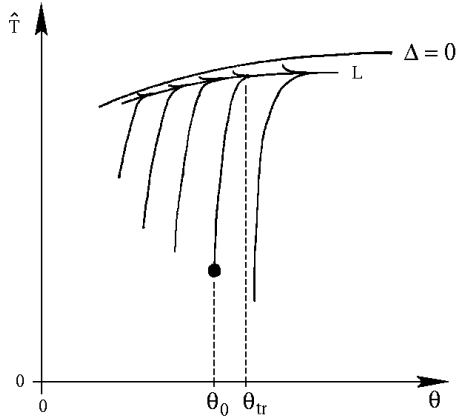}
\caption{Qualitative structure of the field of integral curves.
 We see that the calculations in the direction of increasing $\theta$ are stable 
because of the attraction to the limiting surface $L.$
 The surface $L$ is a separatrix -- it separates the colliding streams of integral curves.
The stream from the $\Delta = 0$ surface repulses the curves
 which come from the "thickness"
 (from the distant region; see the text)
 to this surface.
 Since the ratios $v_{\theta}/c$ and $\sqrt{2T_S/m_p}/c$ are small,
the relative gap
 (for example, $\Delta \hat T/\hat T)$ between the $L$ and $\Delta = 0$ surfaces is small.}
\end{figure}

Thus, any trajectory consists of steep and gentle segments
 (see Fig. 14).
 The steep and gentle segments lie in the distant
 and boundary subregions,
 respectively.
 On the steep segment,
 the first and second of the three terms in Eq. (3.17) are important.
 On the gentle segment, all three terms are comparable.
 On the steep segment, $\Delta$ rapidly decreases with increasing $\theta.$
 As one makes a transition from the steep segment to the gentle one,
 the decrease in $\Delta$ becomes saturated.
 The derivative $|d\ln\Delta/d\theta|$
 is small on the gentle segment,
 implying that the radiative losses are roughly offset by the frictional heating.
 On the steep segment, the energy balance,
 if by the energy balance we mean the equality of the right part of Eq. (3.17) to zero,
 is not maintained --
 the radiative losses are small compared to the heating.
 This implies that the left part of Eq. (3.17) plays an important role here.
 Note that the term in the left part corresponds to the advective transfer
 of radiative enthalpy
 [cf. Subsec. 1.3 (IV)].
 The extent of the steep segment with large (in magnitude) derivatives
 is small:
$\Delta\theta\ll\theta_{\star},$
$\; \Delta\theta = \theta_{tr}-\theta_0,$
 where $\theta_{tr}$ refers to the transition
 from the steep segment to the gentle one on a given trajectory (see Fig. 14).
When the complete system is calculated numerically,
 the short steep segment is traversed in small steps
 (a fraction of $\Delta \theta),$
while the extended gentle segment is traversed with large steps
 (a fraction of $\theta_{\star}).$
 At $\theta \approx \theta_{tr},$
 the trajectory makes a sharp turn.
 As can be seen,
 the trajectories are attracted or caught in the boundary subregion,
 which lies near\footnote{
Accordingly, 
in this part of spread flows,
 the effective or total pressing is small, 
and the layer is thick.}
 the $\Delta = 0$ surface
 and do not leave this subregion.
 For this reason,
 the "levitation" surface is called the attracting or limiting surface,
 hence the stability of the calculations 
and the relative "crudeness" of the results
 with respect to $U_0$ and $\hat T_0$
and, consequently, with respect to the disk parameters\footnote{
Thus, of all the disk parameters,
 only the rate of mass inflow $\dot M$
 is important for the flow in the spread layer.
 The regulation (selection) within the one-parameter $(1d)$ family
 (see below)
 of the flow
 that is actually realized
 is associated with the conditions in the spreading flow
 at the interface between the radiating and dark layers
rather than with the disk parameters.}.

Thus,
 the steep segments along with the distant subregion turn out to be unimportant.
 Only the boundary subregion remains.
 Accordingly, we can immediately take the initial data near the curve (3.16). 
In this way we further narrow down 
(for the third time, see above)
 the set of initial data
$ (\theta_0, U_0, W_0, \hat T_0).$

Let us introduce the polar (angle/radius) coordinates
 in the $(U_0, \hat T_0)$ plane
 instead of the $U_0$ and $\hat T_0$ coordinates.
 Let the angle be related to the ratio
$$
i=\sqrt{\hat T_0}/U_0. \eqno (3.18)
$$
We use the deviation $\delta\Delta$ from the hyperbola (3.16)
instead of the radius.
 One\footnote{
We obtain one value of $i$ instead of the pair of $U_0$ and $\hat T_0.$
 The initial points lie in a narrow band near the hyperbola.}
 parameter $i$ (3.18)
 now runs the set of initial data,
 because the deviation $\delta \Delta$ is insignificant.
 A {\it certain trajectory}
(steady accretion flow)
 corresponds to {\it each} value of $i.$
 We thus arrive at a one-parameter family of solutions,
 which we will call the $1d$ family. 
In phase space,
 the integral curves from this family cover the surface $L$
 near the $\Delta = 0$ surface (see Fig. 14).

\subsection{Within the One-Parameter Family. The Dependence on Latitude.
Critical Regime}

Consider an arbitrary steady flow from the family of solutions
 for a given $\dot M$
 but with different $i.$
 It represents the $\theta-$distribution of $U, W,$ and $\hat T.$
 Let us study this flow.
 There are two characteristic regions above and below $\theta_{\star}.$
 The lower region is the radiating belt.
 In this belt,
$ [W(\theta)]^2 $  is great;
 the loss of rotational energy maintains the radiative flux $q(\theta)$
 at such a level
 that $\Delta \ll 1$
 (or $q \approx q_{eff};$ see Sec. 1).
 The rotational energy is exhausted inside the belt.
 As a result, the luminosity $q(\theta)$ abruptly decreases
 in the transition zone at $\theta \approx \theta_{\star}$ --
the accretion-flow surface darkens.
 Concurrently, the function $\Delta (\theta)$ increases
$[ \Delta (\theta) \approx 1$
 for $\theta > \theta_{\star} ],$
$ g_{wr}$ is large
$ (g_{wr} \approx g_0),$
 and the layer under the strong press contracts
 and becomes very thin outside the radiating belt
$ (h$ is a few centimeters).

The lower (in latitude) region is of greatest interest.
 Since the radiation pressure in this region is large,
 the speed of sound 
$
c_s\simeq\sqrt{p/\rho}
$
 is high compared to the meridional spread velocity
 $v_{\theta}.$
 For this reason, 
the meridional distributions inside the radiating belt become hydrostatic.

In any solution from the family of solutions, the Mach number,
$$
\Ma_{\theta}=\frac{v_{\theta}}{c_s}, \;\;\;\;\;\;\;
c_s^2 = \frac{c^2}{3}\,
\left(1 + \frac{3}{4}\,\frac{\rho_S\, c^2}{a T_S^4}\right)^{-1}
 + \frac{10}{3} \,\frac{T}{m_p},
 \eqno (3.19)
$$
which was calculated for a radiation-dominated plasma (Weinberg 1972),
must increase with $\theta$
 (the pressure decreases, and the flow accelerates).
 Indeed, the quantity (2.11) decreases with increasing $\theta,$
 because its gradient
 counteracts the tangential component of the centrifugal force,
 which tends to return the plasma to the equator 
[see Eq. (2.20)].

Let us compare the $\theta$ dependences for various values of $i$  that run the family.
 The Mach number $\Ma_{\theta}$ (3.19) at the interface 
between the disk and the spread layer 
$ [ \Ma_{\theta} (\theta = 0) ]$
 is smaller,
 if $i$ is larger
 $[i$ is the reciprocal of the Mach number
 without a factor
 $\simeq \sqrt{ G_{eff}/\Delta };$
 cf. the definitions (3.18) and (3.19)].
 We take $\Ma_{\theta} (\theta_0)$
 at the interface between the disk and the spread layer
 and assume that the disk is thin
$ (\theta_0 \ll \theta_{\star}).$
 We can thus write
$ \Ma_{\theta}(\theta = \theta_0) = \Ma_{\theta}(\theta=0).$
 At a fixed $i,$ 
 the function $\Ma_{\theta} (\theta)$
 monotonically increases with $\theta$
 in the interval $0 < \theta < \theta_{\star}.$
 Consequently,
$ \Ma_{\theta} (\theta) $ 
reaches a maximum in this interval at its edge.
 An analysis of the effect of $i$
 shows that
 $\Ma_{\theta} (\theta_{\star}),$
 together with $\Ma_{\theta} (0),$
 increases with decreasing $i.$
 On the $i$ axis,
 the family fills the semiaxis $i_{\star} < i < \infty.$
 The edge $i_{\star}$ corresponds to the critical spread regime,
 in which the {\it sonic point} $\theta_c$
 lies at the boundary of the radiating belt
$[\, \theta_c = \theta_{\star}, \; \Ma_{\theta} (\theta_{\star}) \simeq 1\,].$
$\Ma_{\theta} (\theta)$ also monotonically increases with $\theta$
for $i < i_{\star}.$
 If $i < i_{\star},$
 then the rotational energy at $\theta_c,$
 at which $ \Ma_{\theta} (\theta_c) \simeq 1,$
 has not yet been exhausted:
$ v_{\varphi} (\theta_c) > 0, $
 i.e., the latitude $\theta_c$ 
is inside the radiating belt.
 At the sonic point $\theta_c,$
 the solution becomes two-valued ('tips over').
 Accordingly, we reject the solutions with $i < i_{\star}$ as nonphysical.

{\it The critical regime} occupies a special place.
 It restricts the $1d$ family of subsonic
 (for $\theta < \theta_{\star})$
 flows.
 In this regime, 
the plasma is ejected with the speed of sound from the radiating belt. 
Accordingly,
 the region at $\theta > \theta_{\star}$
 has no effect on the flow in the radiating belt.
 At $i > i_{\star},$
 the regions
$\theta < \theta_{\star}$
 and $\theta > \theta_{\star}$ are coupled.
 For example, the action of lateral forces
 (and friction $\tau_{\theta})$
 between elements $\delta \theta$
 extends from the region $\theta > \theta_{\star}$
 into the region $\theta < \theta_{\star}.$

\section{PHYSICS OF DECELERATION AND SPREAD}

\subsection{Numerical Simulations}

The system (3.12)-(3.14) was integrated numerically.
 In doing so, we approximated the derivatives by finite differences.
 We specified $i$
 to calculate the initial point
 and found the intersection point
with the coordinates $(U_{01}, \hat T_{01})$
 of the curves
$i = i(U, \hat T)$
 and $\Delta (U, \hat T) = 0.$
 We took $U_0 = 0.9 U_{01}$
and $\hat T_0 = 0.9\hat T_{01}$ as the initial values.
 The values of $\theta_0$ and $W_0$ were varied
$ (10^{ - 3} < \theta_0 < 10^{ - 2},$
$  10^{ - 3} < 1-W_0 < 10^{ - 1}).$
The steep segment was traversed with the step
$ \sim 10^{-6} \Delta\theta.$
  We took $\theta, U, W,$ and $\hat T$ at its end as the initial values
 when we calculated the gentle segment with a step
$\sim 10^{-6} \theta_{\star}.$
 The round-off error was $10^{ - 16}$ (double accuracy).
 A preliminary estimate of $\theta_{\star}$
 was obtained from formula (1.1)' for the Eddington scale.

The family consists of subsonic (subcritical)
 and supersonic (supercritical) subfamilies
 in the radiating belt.
 In the solutions from the supercritical subfamily,
 the sonic point lies inside the radiating belt
 (see Subsec. 3.6).
 As was stated, these solutions are discarded.

When we say below that the flow is subsonic,
 we mean that it is subsonic in the radiating belt.
 The determinant $d( \theta )$ (3.15)
 was calculated along an integral curve.
 The first zero of the function $d(\theta),$
 when running in $\theta$ from $\theta_0,$
 is called the sonic point $\theta_c.$
 At this point, the integral curve makes a turn, 
and the solution becomes two-valued.
 Since the determinant $d$ appears in the denominator
 of the derivatives (3.12) and (3.14),
 the derivatives $U'$ and $\hat T'$ have a singularity at $\theta_c.$
 Thus, the solution exists in the interval 
$\theta_0 < \theta < \theta_c.$

\subsection{One-Parameter Family as a Whole
 and the Passage Through the Critical Point}

As was shown in Subsecs. 3.4-3.6,
 the integral curves with general initial data
 form a one-parameter (or $1d)$ family.
 An example of a family is given in Fig. 15
$(\Ma_{\theta}$ profiles, transcritical and supercritical solutions, $100<i<160)$
and Fig. 16
$(\Sigma_S$ profiles in the same family, subcritical and supercritical solutions,
$100 < i < 500).$
 At a fixed $\theta,$ $\;\Ma_{\theta}$ increases,
 while $\Sigma_S$ decreases with decreasing $i.$
 In Fig. 15, the upper and lower solutions refer to $i = 100$ and $160,$
 respectively.
 In Fig. 16,
 the upper and lower solutions refer to $i = 500$ and $100,$
 respectively.
 The supercritical solutions terminate at $\theta_c < \theta_{\star}.$
 The subcritical solutions extend into the region $\theta > \theta_{\star},$
i.e., their termination point $\theta_c$
 lies to the right of $\theta_{\star}.$
 In Figs. 15 and 16, the two upper (Fig. 15) and two lower (Fig. 16) solutions
 are supercritical.
 The arrow in Figs. 15 and 16 marks the critical solution
 with $i = i_{\star}$ and $\theta_c = \theta_{\star}.$
 In Fig. 15,
 the subcritical solutions with $i > i_{\star}$ run below (above in Fig. 16)
 this solution.
 The solutions
 for which $i$ lies in the vicinity of the critical value $i_{\star}$
$\; ( |i - i_{\star}| \ll i_{\star} ).$
 are called transcritical.

\begin{figure}
\epsfxsize=16cm
\epsffile{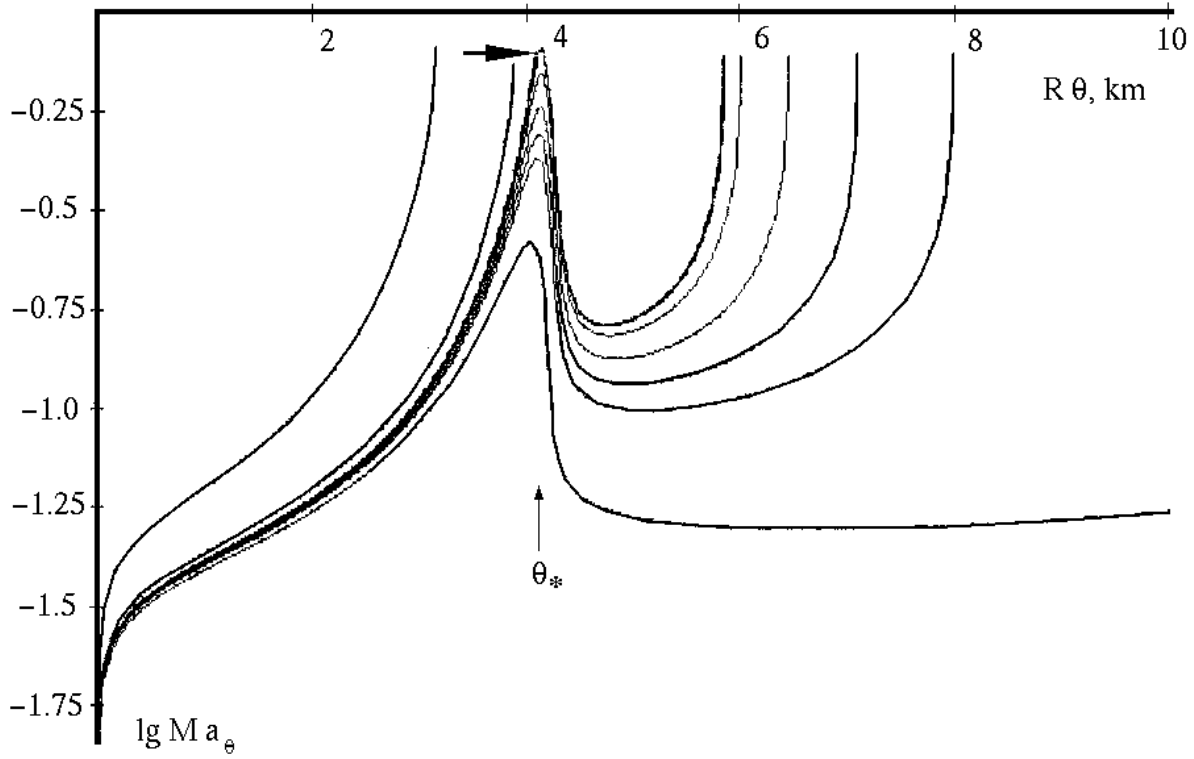}
\caption{One-parameter family of steady spread flows $(100<i<160).$
The dependences $\log [\Ma_{\theta} (R\theta)]$
are for
$\dot M=4\cdot 10^{17}$ g s$^{ - 1},$
$1-W_0=10^{-2},$
$\; \theta_0=10^{-3},$
 and
 $\alpha_b = 10^{-3}.$
$R \theta$ is the arc length along the meridian in km,
$R$ is the stellar radius.
Here $R = 10$ km and $M=1.4 M_{ \odot }.$
 The calculations were performed using Newton's approximation. 
Recall that the curves terminate at the sonic point
$\theta_c,$
 at which the determinant is $d(\theta_c) = 0$
 [see Eq. (3.15) and Subsec. 4.1],
 or are bounded by the computational interval.}
\end{figure}

\begin{figure}
\epsfxsize=16cm
\epsffile{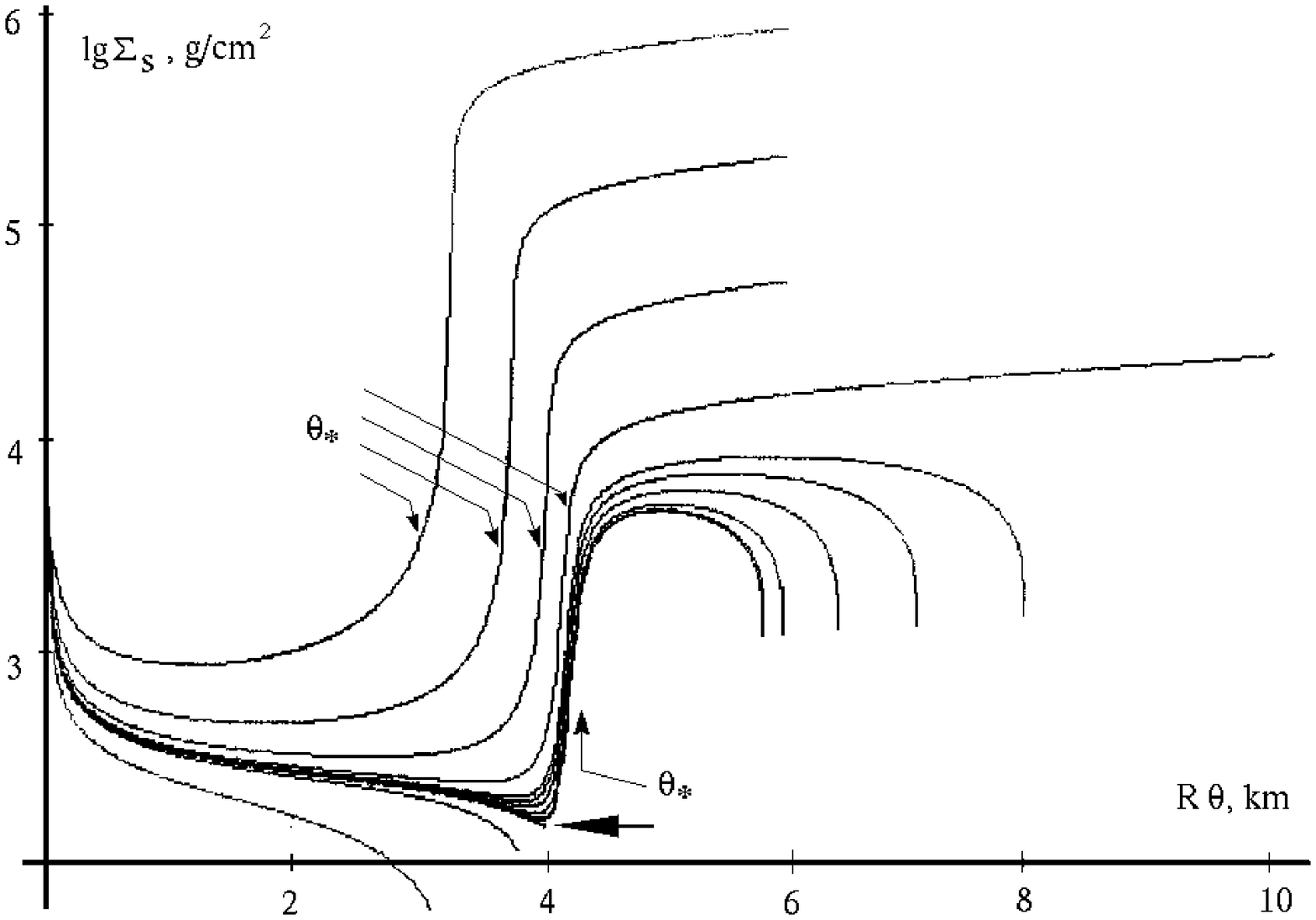}
\caption{$\Sigma_S$ profiles for various values of $i.$
 The four upper profiles rest on the end of the computational interval.
 The remaining profiles terminate at the sonic point.
 The parameters of the variant are given in the caption to Fig. 15.}
\end{figure}

Let us discuss the transcritical behavior.
 The function $\theta_c(i)$ increases
 with increasing $i$ (see Fig. 17 and Figs. 15 and 16).
 This function appears to have a discontinuity at $i_{\star}.$
 In our example,
$ i_{\star} \approx 151.404.$
 At least in our example at the accuracy we chose $(1.5-$mm step),
the discontinuity cannot be reduced by refining $i$
$\; (i_- = 151.403, \;i_+ = 151.405).$
 The discontinuity boundaries in $R \theta$ are:
$R \theta_- = 4.1365$ and $R \theta_+ = 5.8448$ km;
 the relative width is $\Delta \hat \theta =$
$ 2 ( \theta_+ - \theta_- )/( \theta_+ + \theta_- )=0.34.$

\begin{figure}[tbh]
\epsfxsize=8cm
\epsfbox[-200 185 400 656]{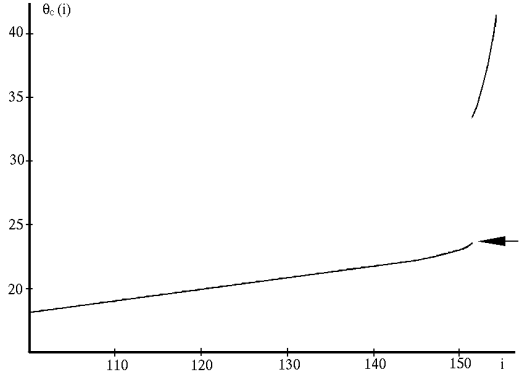}
\caption{Passage of the sonic point $\theta_c$
 through latitude 
$\theta_{\star}$ as $i$ is varied (3.18).
 The latitude $\theta_c$ is given in degrees.
 The parameters of the variant are given in the caption to Fig. 15.}
\end{figure}

The latitude $\theta_{\star}$ divides the flow into two characteristic regions
(Sec. 1 and Subsec. 3.6).
 It separates the rotating (separation in $W),$
 hot (separation in $\hat T),$
 levitating (separation in $\Delta$ and $h),$
 and radiating (separation in $q)$ region
 from the nonrotating, cold, pressed and dark region.
 The separation into these regions is clearly seen
 on the subsonic profiles in Figs. 15 and 16
 (see the arrows $\theta_{\star}).$ 
The transition between them is sharp.
 We can also determine $\theta_{\star}$
from the left boundary of the discontinuity in Figs. 15 and 16
 (see also Fig. 17).
 It is marked by the arrow in Figs. 15-17.
 The discontinuity in $\theta_c(i)$
 is produced by the passage of the sonic point through the belt boundary
 $\theta_{\star}.$

In the momentum equation (3.5),
 the momentum flux in the subsonic flow is small
 compared to the pressure gradient.
 The gradient is balanced either by the component of the centrifugal force $F_{cf}$
 or by the meridional friction $F_{\theta}.$
 Calculations show that the meridional friction\footnote{
Of course, the latitudinal friction $F_{\varphi}$
 in the equations of angular momentum (3.3), (3.6) and energy (3.4), (3.7)
cannot be ignored in the rotating belt.}
$ F_{\theta}$
 and the force $F_{cf}$
 can be disregarded inside and outside the radiating and rotating belt,
respectively.
 This is yet another characteristic
 in which there is a clear separation at $\theta_{\star}.$

{\bf Dark part of the layer. The sonic point and its position.}
In the dark region, the system (3.5)-(3.7) simplifies greatly.
 Here, we may assume that $\Delta = G_{eff} = 1$
 and ignore the rotation of matter.
 The flow as a whole
 (the radiating and dark layers)
 is numerically analyzed below.
 The numerical analysis must be preceded by a qualitative study.
 Since the variability of $\cos\theta$ in the equations 
is of little importance for a qualitative analysis,
 we set $\cos\theta = 1.$
 The qualitative analysis leads us to conclude
 that the flow terminates at the sonic point,
 as confirmed below by the numerical calculations
 which are free from the assumed simplifications.
This conclusion is important in constructing the ensuing theory
 of slow spread and settling of matter over a neutron-star surface,
 because in this way the radiative-frictional mechanism of spread deceleration
is revealed.

As a result of the simplifications above,
 we obtain
$$
(5U^2-4\hat T)U' + 4 U\,\hat T' = -5\alpha_b U^3/\hat T, \eqno (4.1)
$$$$
5 U U' + 14\hat T'= - 5 A\, U\,\hat T^4.  \eqno (4.2)
$$
Resolving the system (4.1), (4.2) for the derivatives,
 we obtain
$$
U' = 10 U^2 (2 A \hat T^5-7\alpha_b U)/\hat T d_1, \eqno (4.3)
$$$$
\hat T' = 5 U [\, (4\hat T-5 U^2)A\hat T^5 + 5\alpha_b U^3]/\hat T d_1, \eqno (4.4)
$$$$
d_1=50 U^2 - 56\hat T. \eqno (4.5)
$$
Eliminating $\theta$ from the system (4.3), (4.4),
 we arrive at the equation
$$
\frac{d\hat T}{dU} =
\frac{(4\hat T-5U^2)(A/\alpha_b)\hat T^5 + 5 U^3 }{
2 U\, [\,2(A/\alpha_b)\hat T^5 - 7 U\,]} \eqno (4.6)
$$
in the phase plane,
 which allows an effective qualitative analysis,
because it is two-dimensional.

Let us estimate the ratio of the frictional
$(\alpha_b U)$
 and radiative 
$(A\hat T^5)$ terms in Eq. (4.6).
 For 
 $\dot M_{17}\sim 1,$
$\; \alpha_b\sim 10^{-3},$
$\; T\simeq 0.5$ keV,
$ U\sim \sqrt{\hat T}$
 they are comparable.
 We offset the large $(\sim 10^{23})$ constant $A/\alpha_b$
 by changing the scale
$$
U\to \sqrt{\lambda} \hat U,\;\;\;\;\;\;\;
\hat T\to\lambda\tilde T,\;\;\;\;\;\;\;
\lambda=(\alpha_b/A)^{2/9}=
2.7\cdot 10^{-5} \,\alpha_b^{2/9} \,\dot M_{17}^{4/9},
$$
[see (3.9) and (3.11)].
 Note that the velocity and temperature at $\hat U = \tilde T = 1$
 are
$v_{\theta}=330 \,\alpha_3^{1/9} \,\dot M_{17}^{2/9}$ km s$^{ - 1}$
 and
$\,T=0.57\,\alpha_3^{2/9} \, \dot M_{17}^{4/9}$ keV,
 where we set $\alpha_b = 10^{ - 3} \alpha_3.$
 Replace $\hat U, \tilde T$ by $\Ma_{\theta}, \tilde T.$
In the dark flow,
 we have
$ \Ma_{\theta}^2 = (3/5) \hat U^2/\tilde T$
 [see (3.19)].
 Making the replacements
$X=\sqrt{10/9} \Ma_{\theta}$
 and
$Y= \tilde T^{9/2},$
 we can smooth out the fourth power of temperature from Stefan-Boltzmann's law,
 which complicates the analysis.
 In these variables,
 Eq. (4.6) takes the form
$$
\frac{dY}{dX}=
9\,\frac{Y}{X}\;
\frac{    \sqrt{2/3} (8-15X^2)Y + 15 X^3 }
{         \sqrt{2/3} (8+15X^2)Y-56X-15X^3}. \eqno (4.7)
$$

A phase portrait of Eq. (4.7) is shown in Fig. 18.
 The function 
$[X]_c (i) = X [\theta_c (i)] = X(i)$
 gives the Mach number at the end of the integral curve (at the sonic point).
 It is constant
$[X (i)\equiv \sqrt{56/75}]$
 at 
$i > i_{\star},$
 i.e., when the sonic point lies in the dark region.
 In the radiating region,
 in which $i < i_{\star}$
 and $\theta_c < \theta_{\star},$
 this function slightly changes with $i.$
 For $100 < i < i_{\star},$
$\;X(i)$ is smaller than $\sqrt{56/75}$ by $5\%$
 and abruptly increases to this value at $i_{\star}.$
 The vertical line 4
$\; (X = \sqrt{56/75})$
 in Fig. 18 is a locus of sonic points $\theta_c.$
 On this line, the determinants $d$ (3.15)
 and $d_1 = \lambda \, \tilde T (75X^2 - 56)\,$ (4.5)
vanish.
 It is easy to show
 that the vertical line 4 and the isoclines of zeros (curve 2)
 and infinities (curve 1) intersect at a single point.

\begin{figure}[tbh]
\epsfxsize=12cm
\epsfbox[-100 207 500 633]{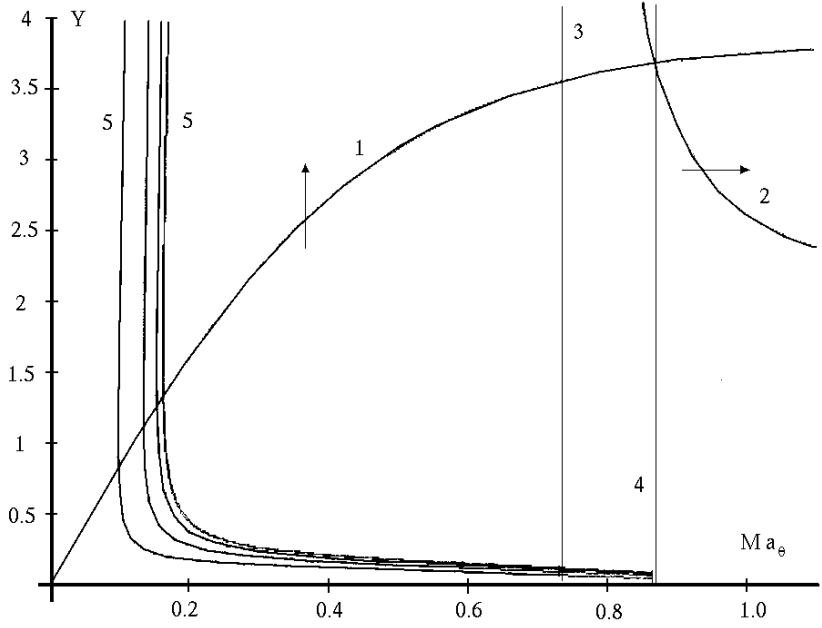}
\caption{Phase plane of Eq. (4.7):
1 -- the isocline of infinities,
the arrows indicate the orientation of integral curves on the isoclines;
2 -- the isocline of zeros; 
3 -- its asymptotic limit for $Y \to \infty,$
4 -- the sonic point
$(d_1 = 0,$ $\;\Ma_{\theta} = \sqrt{56/75}).$
 The integral curves 5
 were taken from the calculations of the complete system 
which are shown in Figs. 15-17
 (the subsonic subfamily,
$i > i_{\star}).$}
\end{figure}

It follows from the qualitative analysis of Eq. (4.7)
 that the integral curves
 which start from the subsonic region
 must necessarily intersect the vertical line 4.
 Indeed, the segment of the integral curve at
$\theta>\theta_{\star}$ 
 starts from the subsonic region.
 The derivative $dY/dX$ (4.7) is positive
 above the isocline of infinities at
$X<\sqrt{8/15},$
 while the derivative $dY/d\theta$ is negative [see Eq. (4.4)].
Hence, on the integral curve in the $X, Y$ plane,
 the motion occurs in the direction of decreasing $Y$ and $X$
 as $\theta$ increases.
 The $Y$ semiaxis cannot be intersected.
 The integral curve 5
 must then necessarily intersect the isocline of infinities
 (see Fig. 18).
 Below the isocline of infinities, $dY/dX$ is negative.
 This time the $X$ semiaxis cannot be intersected.
 The isocline of zeros 2 lies behind the "triangle"
 that is formed by the segments of the $X$ semiaxis,
 the vertical line 4,
 and the isocline of infinities 1.
 Therefore,
 the stream of integral curves 5 arrives at the vertical line 4.
 The curves crowd together near the limiting curve that corresponds to $i_{\star}.$
 The initial value of $X$ (or $\Ma_{\theta})$
 decreases with increasing $i.$
 Thus, 
we established the causes of the "tipping"
 and determined the position 
of the sonic point in the dark region by analyzing the simplified system 
(4.1) and (4.2).

\subsection{Preference of the Critical Regime. The Pedestal Under
 the Radiating Belt} 

In Figs. 1-4
 we assumed that the surface $S$ is equipotential ("horizontal").
 The equipotential surface of a nonrotating star is spherical.
 Let us find out whether the underlying surface can be spherical.
 Calculations show that the surface density $\Sigma_S (\theta)$ of the plasma
 which constitutes the spreading layer is higher in the dark region (Fig. 10).
 In addition, the acceleration 
$g_{eff} = [\,1 - U^2 (\theta) - W^2 (\theta)\,]\, g_0$
 in a differentially rotating radiating belt is smaller
 than that in the dark region.
 Consequently, the weight 
$g_{eff} (\theta)\,\Sigma_S (\theta)$
 of the spreading layer per unit area is larger in the dark region;
 the pressure of the spreading layer on the underlying layers is also higher.
 We ignore the motion in the underlying layer 
beneath the surface $S.$
 Since the motion in this layer is strongly subsonic,
 the hydrostatic contribution to the pressure is more important 
than the dynamical contribution.
 Because of the difference in pressure on the surface $S$
 from the spreading layer under the radiating region, 
this surface rises to form a pedestal.
 Let
$h_{US}(\theta) = R_S(\theta)-R_S(90^0),$
where 
$R_S$ is the radius of this surface.
 We have
$$
(h_{US})_{hs}=\frac{p_d-p_b}{\rho_{US} \; g_0}\simeq
\frac{p_d/\rho_{US}}{g_0}\simeq \frac{(c_s)_{US}^2}{g_0}
\simeq 17\,T_{US},
$$
because there is no rotation under $S$
 and because the total pressing force $g_0$ is in action.
 Here, $p_d$ and $p_b$ are the pressures on the surface $S$
 under the dark and radiating regions, respectively;
$\rho_{US}$ is the density of the matter under $S;$
$ (c_s)_{US}$ and $T_{US}$ are the speed of sound and the temperature
under  $S;$
 and $h_{US}$ and $T_{US}$ are given in cm and keV, respectively.

In addition, the radiating belt is considerably hotter
 with 
$T_{US}(\theta) = T_S(\theta)$ (see Subsec. 2.1).
 As noted above,
 the radiating belt heated up the underlying layers,
 causing the pedestal height to increase by $(h_{US})_T \simeq (c_s)^2_S/g_0.$

Let us compare the height
$(h_{US})_{hs} + (h_{US})_T$
 with the thickness
$$
h_{eff}(\theta)=\frac{1}{| (d\ln \rho/dy)_S \,|}=
\frac{4}{3}\;\frac{R\;\hat T}{\Delta}
$$
of the spreading layer,
(see Fig. 5).
 We see that the thickness of the spreading flow in the radiating belt is much larger 
than the pedestal height.
 This is because $g_{wr}$ and $g_0$ differ. 
Consequently, the relief of the surface $S$ is not important
 for the flow in the radiating belt.

At the same time,
 the height $h_{US}(0)$ is of the order of the thickness
 of the spreading dark layer,
 implying that the dark flow cannot affect the flow in the radiating belt.
 The flow in the belt must therefore be transcritical.

\subsection{Spread of Dark Matter}

The dark flow is specified by the functions $v_{\theta}$ and $T_S.$
 Its segment
$\theta_{\star}<\theta<\theta_c$
 terminates at the sonic point 
(see Subsec. 4.2).
 The separation
$\theta_c - \theta_{\star}$
 increases with a decreasing initial Mach number,
 for example,
$\Ma_{\theta} (\theta_{\star}).$
 In order to reach the pole
$(\theta_c>90^0),$
 it is necessary that
$\Ma_{\theta} (\theta_{\star})$
 be sufficiently small.

In the model (3.2)-(3.4),
 the surface $S$ is impenetrable
 [integral (3.1)].
 Above this surface,
 the newly accreted matter is transported 
from the source at the equator to the sink.
 Let us consider the sink. 
The dark layer is thin compared to the stellar radius,
$h \ll R.$
 It may 
mix with the underlying fluid
 when traversing a long path,
$(s_{mix}-s_{\star})/h \gg 1;$
 the boundary $S$ terminates in the mixing zone.

In another case, the mixing is ineffective,
 and the dark flow runs up to the pole.
 In this case,
 we have a "bath-filling" flow with a filling "tap"
 near the boundary $\theta_{\star}.$
 The mixing zone then is located not far from the tap: 
$s_{mix}-s_{\star}$ 
 is of the order of $(10-100) h.$
 In the "bath",
 the matter is essentially in hydrostatic equilibrium
 and, naturally, moves at velocities much lower than the speed of sound.
 This matter flows under the radiating belt
 (especially since the pressure under the radiating layer
 is lower than the pressure under the dark layer; see Subsec. 4.3).
The boundary $S$ is preserved under this belt.
 The "bath" filling is very slow,
 and the flow is strongly subsonic.
 Accordingly, the weight contribution $(h_{US})_{hs}$ to the pedestal vanishes,
 and only the temperature contribution $(h_{US})_T$ remains.
 Note the possibility of subsonic circulation under $S$
 not only due to the viscous entrainment 
(gripping) by the high-velocity layer
 (see Subsecs. 1 and 2.1)
 but also 
due to the difference in the heights of equal pressure.
 In the Earth's atmosphere,
 this leads to trade-wind circulation
 which is attributable to solar heating of the equatorial zone
 (Palmen and Newton 1969; Bubnov and Golitsyn 1995).
 The filling causes the fluid to rise at the velocity 
$v_a$
 or 
to settle at the same velocity.
 This is the radial component of the total velocity.
 It is easy to see that
$v_a\simeq (1/2) (h/R) (\rho_S/\rho_{US})\, v_{\theta}$
 is small
 compared to the spread velocity $v_{\theta}$ in the radiating belt --
 the matter traverses the belt rapidly but settles slowly.
 At the parameters of the problem under consideration,
$ v_a \sim 10$ m s$^{ - 1}.$
 Such velocity ensures a uniform increase in the surface density
 over the entire stellar surface.
 A particular regime of the motion of the dark fluid
 affects the dynamics of a loose belt only slightly.
 Accordingly, the approach (3.1)-(3.4) is acceptable
 for describing the belt all the way to the pole.
 Of course, 
flows which run up to the pole require a special analysis.
 It is necessary to introduce the parameter $v_a,$
 change the boundary condition downstream, 
and add the equation for $\Sigma_S$ to the system (3.2)-(3.4)
 instead of the integral (3.1).
 Note also
 that the allowance for the moderate penetrability of the surface $S$
 under the radiating and rotating belt
 results in certain quantitative changes,
 although qualitative conclusions remain valid.
 This can be seen from the results (see below)
 on the effect of variations in the accretion rate $\dot M$ on the belt structure.

\subsection{Effect of Friction}

Since the intensity of radiation cooling cannot exceed the Eddington limit,
 the area of the emitting surface must be large enough
 to remove the released heat [see (1.1), (1.1)', (1.2), (1.2)'].
 The system under study has excess energy,
 if the emitting area is larger than the area of the disk base.
 The disk accretion in the disk-layer transition zone
 then gives way to the spread of matter over the stellar surface.
 The flow boundary
( in the plane
 that passes through the polar axis)
 turns $90^0$ in the transition zone
 (see Figs. 1-4 and Sec. 1).
 In this case, 
the radial flow
 (disk, $|v_r| \gg v_{\theta})$
 transforms to the meridional flow
 (spread, $|v_r| \ll  v_{\theta}).$
 Because of the delay in the deceleration,
 a reservoir filled with fluid
 which retains its rotation
carried from the disk
 is formed.
 Its surface emits radiation (radiating belt).

To assess the role of the friction model,
 we varied the coefficient $\alpha_b$ in (2.16).
 The cases with $\alpha_b = 10^{ - 3}$ and $10^{ - 3}/3$ 
are compared in Fig. 19.
 The transcritical profiles,
which refer to the discontinuity boundaries
$i_-$ and $i_+,$
 are shown in this figure
 (see Subsec. 4.2).
 Let us compare the transcritical profiles for these values of $\alpha_b.$
 The critical value of $i_{\star}$
 increases roughly proportional to $1/\alpha_b$
 (by a factor of 2.7).
 The functions 
$\Ma_{\theta} (\theta)$
 and
$ U(\theta)$
 decrease,
 while 
$\Sigma_S(\theta)$ increases
$\propto\alpha_b$
 and
$\propto 1/\alpha_b,$
respectively.
 The value of $\theta_{\star}$ increases by $6\%.$
 The functions
$W(\theta)$
 are virtually equal,
differing by only fractions of a percent. 
Consequently, the $q(\theta)$ profile [see (1.2)] is invariant to $\alpha_b.$
 The bottom temperature $T_S$ rises
$\propto (1/\alpha_b)^{1/4}$
 due to the increase in $\Sigma_S$ [see (2.14)].

\begin{figure}[tbh]
\epsfxsize=10cm
\epsfbox[-180 128 403 750]{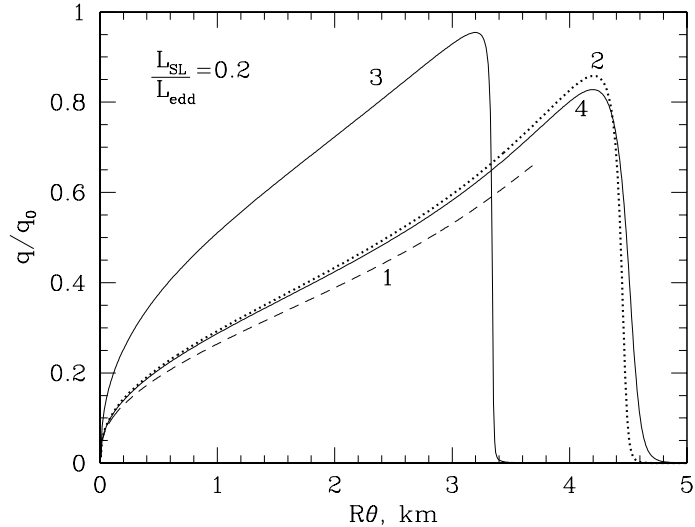}
\caption{Response of the function $q(\theta)$
 to variations in $i$ and $\alpha_b.$
 Curves 1, 2, and 3 refer to moderately supersonic
 (1), slightly subsonic (2), and moderately subsonic (3)
 spread regimes for $\alpha_b = 10^{ - 3.}$
 Curve 4 gives a subsonic transcritical profile 
for $\alpha_b = 10^{ - 3}/3.$
 We see that variations in $\alpha_b$ affect the $q$ profile only slightly.}
\end{figure}

The degree of levitation is
$\Delta \simeq 10^{ - 3}-10^{ - 2},$
since it responds to a variation in $\alpha_b$
 in the same way $T_S$ does.
The height $h_{eff}$
 (see Subsec. 4.3)
 is invariant to $\alpha_b$ variation
with the same accuracy as $W.$

As can be seen,
 there are parameters
 which change significantly as $\alpha_b$ is varied
 (these include $i_{\star},$ $U,$ $\Ma_{\theta},$ $\Sigma_S)$
 and parameters
 which change weakly $(T_S, \Delta, \theta_{\star})$
 or only slightly
 $(h_{eff}, W, q).$
 Note that in view of the behavior noted above, 
the tangential stress (2.16) belongs to the parameters
 which respond only slightly,
 because
$\rho_S=(4/3)\Sigma_S/h_{eff}\propto 1/\alpha_b.$
 This also follows from Eqs. (2.22) and (3.3)
 and from the invariance of $W(\theta).$
 The width of the discontinuity
 in $\theta_c$ at $i_{\star}$ $\;(\Delta \hat\theta,$ see Subsec. 4.2)
 depends on $\dot M$ and $\alpha_b.$

\subsection{Effect of Initial Parameters}

It is clear from the arguments
 which lead to formula (1.2)
 that the equatorial part of the rotating belt radiates weakly,
 because the normal component of the centrifugal force is large here.
 Thus,
 there are two bright belts
 at latitudes $\theta_{\star}$ and $ - \theta_{\star}.$
 The greatest brightness is reached near the edge of the radiating belt.
 The maximum is appreciably lower than unity,
 because
 where $W \simeq 1,$
 the subtraction of the centrifugal component is significant
 and where $W$ is small,
there is nothing to radiate.
 It follows from the smallness of $\Delta$
compared to $G_{eff}$
 that $q \approx q_{eff}$
 [see Sec. 1 for a discussion of formula (1.2)].

Let us analyze the effect of the initial parameters $W_0$ and $\theta_0,$
associated with the disk,
 on the results.
 We compare the cases with $1-W_0 = 10^{ - 2}$ and $10^{ - 1.}$
 They correspond to a change
 in the deviation of the rotation velocity from the Keplerian velocity
 by an order of magnitude.
 For this change in $W_0,$
 the kinetic energy increases by $2\%,$ 
$i_{\star}$ increases by $42\%,$
 the belt width $\theta_{\star}$ increases by $3\%,$
$\Ma_{\theta}$
 inside the belt decreases by $12\%,$
 the surface density $\Sigma_S$ increases by $7\%,$
 the rotation velocity $W$ increases by $1.4\%,$
the temperature rises by $0.8\%,$
 the degree of levitation $\Delta$ decreases by $13\%,$
 the height $h_{eff}$ increases by $14\%,$
 and $q$ decreases by $3\%.$

Let us increase the meridional extent of the disk base by a factor of 10:
$\theta_0=10^{-3}$
 is replaced by $10^{ - 2.}$
 The critical value of $i_{\star}$
 increases by $8\%.$
 The distributions of
$\Ma_{\theta}, \Sigma_S, W$
and other variables for disks of thickness
$2 R \theta_0 = 20$ and 200 m differ only slightly.
 The distribution in $\theta$ seems to shift
 by an amount of the order of the difference between the values of $\theta_0.$
 Thus, for $1-W_0 \ll 1$ and $\theta_0 \ll \theta_{\star},$
variations in $W_0$ and $\theta_0$ have a marginal effect on the results,
in particular, on the $q (\theta)$ profile.

\subsection{Dependence on Accretion Rate}

Above, we discussed the main points 
regarding variations in $\alpha_b,$
$W_0,$
 and $\theta_0.$
 Let us now study the effect of $\dot M.$
 The critical value $i_{\star}(\dot M)$
 and the belt width $\theta_{\star}(\dot M)$
 are plotted versus the accretion rate in Fig. 20.
 In the calculations,
 we approached the critical value by compressing the transcritical vicinity
 by the fork method (see Subsecs. 3.6, 4.2, 4.3).
 The width
$\theta_{\star}(\dot M)$
 was determined from the last calculated supercritical profile.
 We took the position of the sonic point in this profile as the belt width.
 An analysis of the effect of variations in $\dot M$
 on the spread-flow structure 
shows that the relative discontinuity width
$\Delta\hat \theta$ (Subsec. 4.2)
 decreases with increasing $\dot M.$
 At $\dot M \approx \dot M_{pole},$
 the discontinuity disappears,
 and the dependence on $i$ becomes smooth;
 the edge of the radiating belt comes close to the pole.
 Under these conditions, the edge of the rotating belt
 was taken as
$\theta_{\star}(\dot M).$
 We chose the smallest value of $i$
 at which this edge lies at a distance from the pole
 and took it as $i_{\star}.$

\begin{figure}[tbh]
\epsfxsize=9cm
\epsfbox[-180 128 403 750]{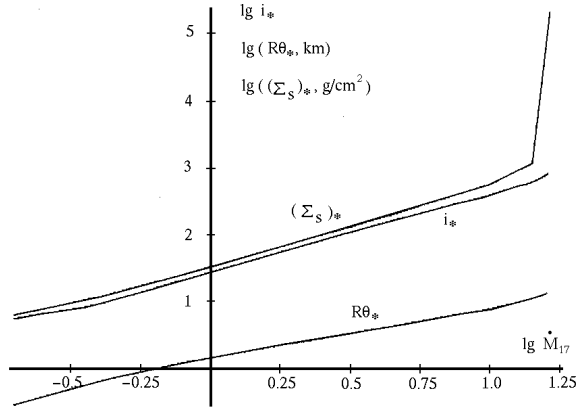}
\caption{Critical value of $i_{\star},$
 radiating-belt width
$R \theta_{\star},$
 and surface density $(\Sigma_S)_{\star}$
 at the belt boundary versus $\dot M.$}
\end{figure}

In Fig. 20,
 we also plotted $(\Sigma_S)_{\star}$ versus $\dot M.$
 The value of $(\Sigma_S)_{\star}$
 is
$\Sigma_S (\theta_{\star}),$ 
 where $\theta_{\star}$ was determined from the last calculated supercritical profile.
 Near the point of stoppage at the pole
 for $\dot M \approx M_{pole},$
 we took the edge of the rotating belt as
$\theta_{\star}.$
 The function
$(\Sigma_S)_{\star} (\dot M)$
 begins to rapidly increase near the pole.
 It is physically clear
 that after the point of stoppage of the loose radiating belt at the pole,
 the spread of matter slows down greatly (Subsec. 4.4).
 The slowdown causes the surface density $\Sigma_S$ to increase proportionally.
 The power law fitting the dependences are
$i_{\star} \approx (\Sigma_S)_{\star}$ [g cm$^{ - 3}]$ 
$\approx 35 \dot M_{17}^{1.14},$
$ \; R \theta_{\star} = s_{\star}$ [km] $\approx 1.4 \dot M_{17}^{0.8}.$

Let us define the contrast
 as the ratio of $\Sigma_S$ in the central zone
 to
$(\Sigma_S)_{\star}.$
 The contrast increases with increasing $\dot M.$
 The fact that it is greater than unity
 is attributable to the existence of a strongly subsonic central zone.
 For this reason,
 the spreading layer stretches when it approaches the edge of the radiating belt.
 While at large $\dot M$ there are a subsonic center and a transonic edge,
 at small $\dot M$
the entire radiating belt is transonic.

\subsection{Limiting Magnetic Field}

Let us estimate the magnetic-field strength
 that still has no effect on the pattern of deceleration and spread outlined above.
 Compare the magnetic pressure
$H^2/8\pi$
 and the pressure $p_S = g_{eff}\,\Sigma_S$ (2.2)
 at the bottom of the spread layer.
 To obtain an estimate,
 we assume that $g_{eff} \sim g_0.$
 We take
$(\Sigma_S)_{\star}$
 as a typical value of
$\Sigma_S$
 and use the fitting dependence on $\dot M$
 which we derived in Subsec. 4.7.
 As a result,
 we obtain $H < H_{max},$
$\; H_{max} = 4\times 10^8 (\dot M_{17})^{0.57}$ gauss.
 A weaker field appears to "get buried" under the layers of accreting plasma.
 X-ray bursters give an example of neutron stars with a weak magnetic field.

\section*{CONCLUSION}

The spread of matter
 during disk accretion was considered.
 We found a radiating belt to be formed in a certain range of $\dot M.$
 The belt width 
$\theta_{\star}$
 depends on $\dot M.$
 At
 $L_{SL}/L_{edd}\simeq 3 \times 10^{-3},$
 the width 
$\theta_{\star}$
is of the order of the thin-disk thickness
$\theta_0=10^{-3}\div 10^{-2}.$
 The belt disappears at
$L_{SL}/L_{edd} \approx L_{pole}/L_{edd} \approx 0.9,$
 because the entire stellar surface radiates.
 The belt rotates.
 Its rotation velocity depends on latitude.
 This belt levitates in the sense
 that the pressing acceleration $g_{wr}$ is small
 compared to the gravitational acceleration $g_0.$
 A delay in the deceleration of rotation is responsible for the appearance of the belt.
 This delay in the deceleration is attributable to a great decrease in the bottom density 
due to the levitation and loosening of the matter.

\section*{ACKNOWLEDGMENTS}

We wish to thank G.S. Golitsyn and N.I. Shakura for helpful discussions.
We also wish to thank N.R. Sibgatullin, a referee of the paper, 
for a careful reading of the manuscript,
 for the checking of the equations, 
and for valuable remarks.

\section*{REFERENCES}

\noindent Barret, D., Bouchet, L., Mandrou, P., et al., Astrophys. J., 1992, vol. 394, p. 615.

\noindent Bisnovatyi-Kogan, G.S., Mon. Not. R. Astron. Soc., 1994, vol. 269, p. 557.

\noindent Boubnov, B.M. and Golitsin, G.S.,
Convection in Rotating Fluids,
Fluid Mech. and its Appl., vol. 29,
Dordrecht: Kluwer, 1995.

\noindent Campana, S., Stella, L., Mereghetti, S., et al., astro-ph/9803303, 26 Mar, 1998a.

\noindent Campana, S., Colpi, M., Mereghetti, S., et al., astro-ph/9805079, 6 May, 1998b.

\noindent Chandrasekhar, S., Radiative Transfer, New York: Dover, 1960.

\noindent Chugaev, R.R., Gidravlika (Hydralics), Leningrad: Energoizdat, 1982.

\noindent Felten, J.E. and Rees, M.J., Astron. Astrophys., 1972, vol. 17, p. 226.

\noindent Gilfanov, M., Revnivtsev, M., Sunyaev, R., and Churazov, E., Astron. Astrophys., 1998,
 vol. 338, p. L83.

\noindent Illarionov, A.F. and Sunyaev, R.A., Astrophys. Space Sci., 1972, vol. 19, p. 61.

\noindent Illarionov, A.F. and Sunyaev, R.A., Soviet Astron., 1975,
 vol. 18(4), p. 413.

\noindent Landau, L.D. and Lifshitz, E.M., Gidrodinamika (Hydrodynamics), Moscow: Nauka, 1986.

\noindent Lapidus, I.I. and Sunyaev, R.A., Mon. Not. R. Astron. Soc., 1985, v. 217, p. 291.

\noindent Lin, C.-L., Moeng, C.-H., and Sullivan, P.P., Phys. Fluids, 1997, vol. 9(11), p. 3235.

\noindent Loitsyanskii, L.G., Mekhanika zhidkosti i gaza (Fluid and Gas Mechanics),
Moscow: Nauka, 1973.

\noindent Lynden-Bell, D. and Pringle, J.E., Mon. Not. R. Astron. Soc., 1974, vol. 168, p. 603.

\noindent Meyer, F. and Meyer-Hofmeister, E., Astron. Astrophys., 1989, vol. 221, p. 36.

\noindent Narayan, R. and Yi I., Astrophys. J., 1995, vol. 452, p. 710.

\noindent Palmen, E. and Newton, W., Atmospheric Circulation Systems,
N.Y. and London: Academic Press, 1969.

\noindent Papaloizou, J.C.B. and Stanley, G.Q.G., Mon. Not. R. Astron. Soc., 1986, vol. 220, p. 593.

\noindent Popham, R. and Narayan, R., Astrophys. J., 1995, vol. 442, p. 337.

\noindent Popham, R., Narayan, R., Hartmann, L., and Kenyon, S., Astrophys. J., 1993, vol. 415,
 p. L127.

\noindent Pringle, J.E. and Savonije, G.J., Mon. Not. R. Astron. Soc., 1979, vol. 187, p. 777.

\noindent Schlichting, H., Grenzschicht--Theorie, Karlsruhe: G. Braun, 1965.

\noindent Shakura, N.I. and Sunyaev, R.A., Adv. Space Res., 1988, vol. 8, p. 135.

\noindent Shakura, N.I. and Sunyaev, R.A., Astron. Astrophys., 1973, vol. 24, p. 337.

\noindent Shakura, N.I., Soviet Astron, 1972, vol. 16(3), p. 532.

\noindent Sibgatullin, R.N. and Sunyaev, R.A., Astron. Lett., 1998,
 vol. 24(6), p. 774. (Astro-ph/9811028).

\noindent Sobolev, V.V., Uchenye Zapiski Leningr. Univ., Ser. Matem. Nauk, 1949, vol. 18, issue 116, p. 3.

\noindent Spitzer, L., Physics of Fully Ionized Gases, N.Y.: Interscience,
 1962.

\noindent Sunyaev, G.A. and Shakura, N.I., Soviet Astron. Lett., 1986,
 vol. 12(2), p. 117.

\noindent Sunyaev, R.A. and Titarchuk, L.G., Astron. Astrophys., 1980, vol. 86, p. 121.

\noindent Tylenda, R., Acta Astron., 1981, vol. 31, p. 267.

\noindent Van der Klis, M., astro-ph/9812395, 22 Dec, 1998.

\noindent Weinberg, S., Gravitation and Cosmology: Principles and Applications of 
the General Theory of Relativity, N.Y. etc.: John Wiley and Sons, 1972.

\end{document}